\documentclass[a4paper,12pt]{article}
\usepackage{amsmath}
\usepackage{pstricks}
\usepackage{pst-node}
\usepackage[ansinew]{inputenc}
\usepackage{amssymb,amsmath}
\usepackage{amsfonts}
\usepackage{epsfig}
\usepackage[english]{babel}
\usepackage{graphics}
\usepackage{epic}
\usepackage{eepic}

\usepackage{hyperref}

\def\p{\partial}

\def\a{\alpha}
\def\b{\beta}
\def\g{\gamma}

\def\o{\omega}

\def\e{\varepsilon}

\font\Sets=msbm10
\def\Integer {\hbox{\Sets Z}}

\def\be{\begin{equation}}
\def\ba{\begin{array}}
\def\ee{\end{equation}}
\def\ea{\end{array}}
\def\bea {\begin{eqnarray}}
\def\eea {\end{eqnarray}}
\def\bean{\begin{eqnarray*}}
\def\eean{\end{eqnarray*}}

\def\RA {\ \Rightarrow\ }

\begin{document}

\title{Symbolic Computation for \\ Nonlinear Wave Resonances}
\author{ E. Kartashova, C. Raab, Ch. Feurer, \\
G. Mayrhofer, W. Schreiner \\
Research Institute for Symbolic Computation \href{http://www.risc.uni-linz.ac.at/}{(RISC)} \\
Johannes Kepler University Linz \\
Altenbergerstr.~69, A-4040 Linz, Austria\\ \\
}


\maketitle
\newpage
\tableofcontents
\newpage
\setlength{\parindent}{0pt}
\setlength{\parskip}{\medskipamount}


\section{Introduction}
Resonance is a common thread which runs through almost every branch
of physics, without resonance we wouldn't have radio, television,
music, etc. Resonance causes an object to oscillate, sometimes the
oscillation is easy to see (vibration in a guitar string), but
sometimes this is impossible without measuring instruments
(electrons in an electrical circuit). A well-known example with
Tacoma Narrows Bridge (at the time it opened for traffic in 1940, it
was the third longest suspension bridge in the world) shows how
disastrous resonances can be: on the morning of November 7, 1940,
the four month old Tacoma Narrows Bridge began to oscillate
dangerously up and down, tore itself apart and collapsed. Though
designed for winds of 120 mph, a wind of only 42 mph caused it to
collapse. The experts did agree that somehow the wind caused the
bridge to resonate, and nowadays, wind tunnel testing of bridge
designs is mandatory.

Another famous example are the experiments of Tesla  who studied in
1898 experimentally vibrations of an iron column and noticed that at
certain frequencies specific pieces of equipment in the room would
start to jiggle. Playing with the frequency he was able to move the
jiggle to another part of the room. Completely fascinated with these
findings, he forgot that the column ran downward into the foundation
of the building, and the vibrations were being transmitted all over
Manhattan. The experiments had started sort of a small earthquake in
his neighborhood with smashed windows, swayed buildings, and panicky
people in the streets. For Tesla, the first hint of trouble came
when the walls and floor began to heave \cite{Tesla}. He stopped the
experiment as soon as he saw police rushing through the door.

The difference between resonances in a human made system and in some
natural phenomena is very simple. We can change the form of a bridge
and stop the experiment by switching off electricity but we can not
change the direction of the wind, the form of the Earth atmosphere
or the sizes of an ocean. What we can try to do is {\it to predict}
drastic behavior of a real physical system by computing its
resonances. While linear resonances in different physical systems
are comparatively well studied, to compute characteristics of
nonlinear resonances and to predict their properties is quite a
nontrivial problem, even in the one-dimensional case. Thus, the
notorious Fermi-Pasta-Ulam numerical experiments with a nonlinear
{\bf 1D}-string (carried out more then 50 years ago) are still not
fully understood \cite{FPU}. On the other hand, nonlinear
 wave resonances in continuous {\bf 2D}-media like ocean, space, atmosphere,
 plasma, etc. are well studied in the frame of wave turbulence
 theory \cite{lvov} and provide a sound basis for qualitative and sometimes also
 quantitative analysis of corresponding physical systems. The notion
 of nonlinear wave interactions is crucial in the wave turbulence theory \cite{dias}.
 Excluding resonances allows to describe a nonlinear wave
 system statistically, by wave kinetic equations and power-law energy
 spectra of turbulence \cite{zak2}, and to observe this behavior in numerical experiments  \cite{zak3}.
  Direct computations with Euler
 equations (modified for gravity water waves, \cite{zak4}) show that
 the
 existence of
 resonances in a wave system yield some additional effects
 which are not covered by the statistical description. The role of resonances in the evolution of water wave turbulent systems
 has been studied profoundly by a great number of researchers. One of the most important
 conclusions (for gravity water waves)
made recently in \cite{T07} is the following: "{\it The four-wave
resonant interactions control the evolution of the spectrum at every
instant of time, whereas non-resonant interactions do not make any
significant
 contribution even in a short-term evolution}."

The behavior of a resonant wave system can be briefly described
\cite{AMS} as follows: 1) not all waves take part in resonant
interactions, 2) resonantly interacting waves form a few independent
small wave clusters, such that there is no energy flow between these
clusters, 3) including some small but non-zero resonance width into
consideration {\it does not} destroy the clusters. A model of
laminated wave turbulence \cite{K06-1} allows to describe
statistical and resonant
 regimes simultaneously while methods to compute
resonances numerically are presented in \cite{comp-all1} (idea) and
in \cite{comp-all234} (implementation).
 Our main purpose here is to study the
possibilities of a symbolic implementation of these general
algorithms using the computer algebra system Mathematica.

The implemented software can be executed with local installations of
Mathematica and the corresponding method libraries; however, we have
also developed a Web interface that allows to run the methods from
any computer in the Internet via a conventional Web browser. The
implementation strategy is simple and based on generally available
technologies; it can serve as a blueprint for other mathematical
software with similar features.

We take as our principal example the barotropic vorticity equation
in a rectangular domain with zero boundary conditions which
describes oceanic planetary waves, and show how : (a) to compute
interaction coefficients of corresponding dynamical systems, (b) to
solve resonant conditions, (c) to construct the topological
structure of the solution set, and (d) to use the software via a Web
interface over the Internet.  A short discussion concludes the
paper.

\section{Mathematical Background}

Wave turbulence takes place in physical systems with nonlinear
dispersive waves thatare  described by evolutionary dispersive
NPDEs. The role of the evolutionary dispersive NPDEs in the
theoretical physics is so important that the notion of dispersion is
used for {\it a physical} classification of PDEs into dispersive and
non-dispersive. The well-known  mathematical classification of PDEs
into elliptic, parabolic and hyperbolic equations is based on the
form of equations and can be applied to the second order PDEs on an
arbitrary number of variables. On the other hand,  the physical
classification is based on {\it the form of solutions} and can be
applied to PDEs of arbitrary order and arbitrary number of
variables. In order to construct the physical classification of
PDEs, two preliminary steps are to be made: 1) to divide all
variables into two groups - time- and space-like variables ($ t \ $
and $\ x \ $ correspondingly); and 2) to check that the {\it linear
part} of the PDE under consideration
 has a wave-like solution in the form of Fourier harmonic
$$
\psi (x,t)= A \exp {i[ kx - \o t]}
$$
with amplitude $A$, wave-number $k$ and wave frequency $\o$. The
direct substitution of this solution into the linear PDE shows then
that $\o$ is an explicit function on $k,$ for instance:
$$
\psi_t+\psi_x+\psi_{xxx}=0 \quad \RA  \o ( k)=k - 5 k^3.
$$
If $\ \o \ $ as a function on $\ k\ $ is real-valued and such that $
\ {\rm d}^2 \o /{\rm d} k^2  \neq 0, \ $  it is called {\it a
dispersion function} and the corresponding PDE is called
evolutionary dispersive PDE. If the dimension of the space variable
$\ x\ $ is more that 1, i.e. $\ \vec{x}=(x_1,...,x_p),\ $ $\
\vec{k}\ $ is called the wave-vector and the dispersion function $\
\o=\o(\vec{k})\ $ depends on the coordinates of the wave-vector.
This classification {\it is not complementary} to a standard
mathematical one. For instance, though hyperbolic PDEs normally do
not have dispersive wave solutions, the hyperbolic equation
$\psi_{tt}- \a^2 \psi_{xx} - \b^2 \psi =0 $ has them.

In the huge amount of application areas of NPDEs (classical and
quantum physics, chemistry, medicine, sociology, etc.) a nonlinear
term of the corresponding NPDE can be regarded as small. This is
symbolically written as
\begin{equation}\label{eq}
L(\psi)=-\e N(\psi)
\end{equation}
where $L$  and $N$ are linear and nonlinear parts of the equation
correspondingly and $\e$ is a small parameter defined explicitly by
the  physical problem setting. It can be shown that in this case the
solution $ \psi  $ of (\ref{eq}) can be constructed as a combination
of the Fourier harmonics with amplitudes $ A $ depending on the time
variable and possessing two properties formulated  here  for the
case of quadratic nonlinearity:
\begin{itemize}
\item{} {\bf P1} The amplitudes of the Fourier harmonics satisfy the following system of nonlinear
ordinary differential equations (ODEs) written for simplicity in the real form
\begin{eqnarray}\label{amplitudes}
\dot{A}_1&=&\a_1A_2A_3\nonumber\\
\dot{A}_2&=&\a_2A_1A_3\label{dynsys}\\
\dot{A}_3&=&\a_3A_1A_2\nonumber
\end{eqnarray}
with coefficients $\ \a_i \ $ being functions on wave-numbers;
\item{} {\bf P2} The dispersion function and wave-numbers satisfy the {\it resonance conditions}
\begin{eqnarray}\label{res}
\begin{cases}
\omega (\vec k_1) \pm \omega (\vec k_2)\pm\omega (\vec k_{3}) = 0,\\
\vec k_1 \pm \vec k_2 \pm \vec k_{3} = 0.
\end{cases}
\end{eqnarray}
\end{itemize}
The transition form (\ref{eq}) to (\ref{amplitudes}) can be
performed by some standard methods (for instance, multi-scale method
\cite{na})
 which also yields the explicit form of resonance conditions.

Keeping in the mind our main problem - to find a solution of
(\ref{eq}) - one has to take care of the initial and boundary
conditions. This is done in the following way: the case of periodic
or zero boundary conditions yields {\it integer wave numbers},
otherwise they are real. Correspondingly, one has to find all
integer (or real) solutions of (\ref{res}), substitute corresponding
wave-numbers into the coefficients $\ \a_i\  $ and then look for the
solutions of (\ref{amplitudes}) with given initial conditions.

One can see immediately a big problem which appears as soon as one has
to solve a NPDE with periodical or zero boundary conditions. Indeed,
dispersion functions take different forms, for instance,
$$
\o^2= k^3, \ \  \o^2=k^3+\a k, \ \ \o^2= k, \ \ \o=\a/k, \ \ \o=m/n(n+1) \ \cdots ,
\mbox{etc.}
$$
with $\vec{k}=(m,n),$ $k=\sqrt{m^2+n^2}$ and $\a$ being a constant.
This means that (\ref{res}) corresponds to a system of Diophantine
equations of many variables, normally 6 to 9, with cumulative
degrees 10 to 16. Those have to be solved usually for the integers
of the order $\ \sim 10^3,\ $ which means that computations has to
be performed with integers of order $10^{48}$ and more. Original
algorithms to solve these systems of equations have been developed
based on some profound results of number theory \cite{comp-all1} and
implemented numerically \cite{comp-all234}.

Further on, an evolutionary dispersive NPDE with periodic or zero
boundary conditions is called {\it 3-term mesoscopic system} if it
has a solution of the form
$$
\tilde{\psi}=\sum_{i=1}^s\psi_i (\vec{x},t)= \sum_{i=1}^s A_i \exp {i[ \vec{k}_i\vec{x}_i - \o t]}
$$
and there exists at least one triple $\ \{A_{i_1}, A_{i_2}, A_{i_2}\} \in \{A_i\} \ $
such that {\bf P1} and {\bf P2} keep true with some nonzero
coefficients $\ \a_i, \  \a_i \neq 0 \ \forall i=1,2,3.$

\section{Equations for Wave Amplitudes}
\label{sec:Amplitudes}

\subsection{Method Description}
The barotropic vorticity equation  describing ocean planetary waves
has the form \cite{rectangular}
\begin{equation}\label{BVE}
\frac{\p \triangle \psi}{\p t} + \b \frac{\p \psi}{\p x}= -\e J(\psi, \triangle \psi)
\end{equation}
with boundary conditions $$\psi=0 \quad \mbox{for} \ \ x=0,L_x; \ \
y=0, L_y.$$ Here $\ \b\ $ is a constant called Rossby number, $\ \e
\ $ is a small parameter and the Jacobean has the standard form
$$
J(a,b)=\frac{\p a}{\p x}\frac{\p b}{\p y}-\frac{\p a}{\p y}\frac{\p
b}{\p x}.
$$ First we
give a basic introduction on how a PDE can be turned into a system
of ODEs by a multi-scale method. Using operator notation, our
problem (\ref{BVE}) is viewed as a perturbed version of the linear
PDE $L(\psi)=0$. We pick a solution of this equation, say $\psi_0$,
which is a superposition of several waves $\varphi_j$, i.e.
$\psi_0=\sum_{j=1}^s A_j\varphi_j$, each being a solution itself. To
construct a solution of the original problem we make the amplitudes
time-dependent. As the size of the nonlinearity in (\ref{eq}) is
just of order $\e,$ the amplitudes will vary only on time-scales
$1/\e$ times slower than the waves. Hence we define an additional
time-variable $t_1:=t\e$ called "slow time" to handle this time
scale. So we look for approximate solutions of (\ref{eq}) that have
the following form
$$\psi_0(t,t_1,\vec{x})=\sum_{j=1}^s A_j(t_1)\varphi_j(\vec{x},t)$$
which for $\e=0$ is an exact solution. The exact solution of the
equation is written as power series in $\e$ around $\psi_0$, i.e.
$\psi=\sum_{k=0}^\infty\psi_k\e^k$. In our computation it is
truncated up to maximal order $m$ which in our case is   $m=1$, i.e.
$$\psi(t,t_1,\vec{x})=\psi_0(t,t_1,\vec{x})+\psi_1(t,t_1,\vec{x})\e.$$
Plugging $\psi(t,t_1,\vec{x})$ one has to keep in mind that, since
$t_1=\e t$, we now have $\frac{d}{dt}=\frac{\partial}{\partial
t}+\e\frac{\partial}{\partial t_1}$ due to the chain rule. Equations
are formed by comparing the coefficients of $\e^k$. For $k=0$ this
gives back the linear equation, but we keep the equation for $k=1$.
In particular, for (\ref{BVE}) we arrive at
\begin{eqnarray*}
\frac{\partial\triangle\psi_0}{\partial t}+\beta\frac{\partial\psi_0}{\partial x}&=&0,\\
\frac{\partial\triangle\psi_0}{\partial t_1}+\frac{\partial\triangle\psi_1}{\partial t}+\beta\frac{\partial\psi_1}{\partial x}&=&-J(\psi_0,\triangle\psi_0.)
\end{eqnarray*}
In order to (\ref{amplitudes}), we have to get rid of all other variables. This is done
by integrating against the $\varphi_j$'s, i.e. $\langle.,\varphi_j\rangle_{L^2(\Omega)}$,
and averaging over (fast) time, i.e. $\lim_{T\to\infty}\frac{1}{T}\int_0^T .\,dt$.

\subsection{The Implementation}

 This method was implemented in Mathematica with order $m=1$ in mind
only. So it won't be immediately applicable to higher orders without
some (minor) adjustments. The ODEs are constructed done by the
function
\begin{quote}
\ttfamily
 ODESystem[L($\psi$), N($\psi$), $\psi$, \\
\strut~~~\{x$_1$,..,x$_n$\}, t, domain, jacobian, m, s, A, linwav, \\
\strut~~~\{$\lambda_1$,..,$\lambda_p$\}, paramvalues].
\end{quote}
Basically this function takes the problem together with the solution
of the linear equation as input and computes the list of ODEs for
the amplitudes as output. Its arguments are in more detail:
\begin{itemize}
\item {\tt L($\psi$), N($\psi$)}: Linear and nonlinear part of equation (\ref{eq}), each applied to a symbolic function parameter.
Derivatives have to be specified with {\tt Dt} instead of {\tt D} and the nonlinear part has to be a polynomial in the derivatives
of the function.
\item {\tt $\psi$}: symbol used for function in {\tt L($\psi$), N($\psi$)}
\item {\tt \{x$_1$,...,x$_n$\}, t}: list of symbols used for space-variables, and symbol for time-variable
\item {\tt domain}: The domain on which the equation is considered has to be specified
in the form {\tt \{\{x$_1$,minx$_1$,maxx$_1$\}, ..., \{x$_n$,minx$_n$,maxx$_n$\}\}},
where the bounds on {\tt x$_i$} may depend on {\tt x$_1$,...,x$_{i-1}$} only.
\item {\tt jacobian}: For integration the (determinant of the) Jacobian must also to be
passed to the function. This is needed in case the physical domain does not coincide with
the domain of the variables above, it can be set to {\tt 1} otherwise.
\item {\tt m, s}: maximal power of $\e$ and number of waves considered
\item {\tt A}: symbol used for amplitudes
\item {\tt linwav}: General wave of the linear equation is assumed to have separated
variables, i.e. $\varphi(\vec{x},t)=B_1(x_1){\cdot}...{\cdot}B_n(x_n)\exp(i\theta(x_1,...,x_n,t))$,
and has to be given in the form

{\tt \{B$_1$(x$_1$), ..., B$_n$(x$_n$),
$\theta$(x$_1$,...,x$_n$,t)\}}.
\item {\tt \{$\lambda_1$,...,$\lambda_p$\}}: list of symbols of parameters the functions in {\tt linwav}
depend on
\item {\tt paramvalues}: For each of the {\tt s} waves explicit values of the parameters
\{$\lambda_1$,...,$\lambda_p$\} have to be passed as a list of {\tt s} vectors of parameter values.
\end{itemize}

\begin{verbatim}
ODESystem[linearpart_,nonlinearpart_,fun_Symbol,vars_List,
    t_Symbol,domain_List,jacobian_,ord_Integer,num_Integer,
    A_Symbol,linwav_List,params_List,paramvalues_List] :=
  Module[{B,theta,eq,k},
    eq = PerturbationEqns[linearpart,nonlinearpart,
                          fun,vars,t,ord];
    eq = PlugInGenericWaveTuple[eq,fun,vars,t,A,B,theta,num]
             /. fun[1]->(0&);
    eq = Table[Resonance2[eq,linwav,vars,t,params,A,B,theta,
                   num,paramvalues,k],
               {k,num}];
    Map[Integrate[Simplify[#,And@@(Function[B,B[[2]]<B[[1]]<
                      B[[3]]]/@domain)]*jacobian,
                  Sequence@@domain]&,
        eq,{2}]
    ]
\end{verbatim}

Internally this function is divided into three subroutines briefly
described below.

\subsubsection{Perturbation Equations, General Form}

The first of the subroutines is

{\tt PerturbationEqns[L($\psi$), N($\psi$), $\psi$,
\{x$_1$,...,x$_n$\}, t, m]}.

As mentioned before we approximate the solution of our problem by a
polynomial of degree $m$ in $\e$. This subroutine works for
arbitrary $m$. In the first step we construct equations by
coefficient comparison. Additional time-variables will be created
automatically and  labeled {\tt t[1],...,t[m]}. The output is a list
of $m+1$ equations corresponding to the powers $\e^0,...,\e^m$. The
implementation is quite straightforward. First set
$\psi=\sum_{k=0}^m\psi_k(t,t_1,...,t_m,x_1,..,x_n)\e^k$ in
(\ref{eq}), where $t_k=\e^kt$, i.e.
$\frac{d}{dt}=\frac{\partial}{\partial
t}+\sum_{k=1}^m\e^k\frac{\partial}{\partial t_k}$. Then extract the
coefficients of $\e^0,...,\e^m$ on both sides and assemble the
equations. Finally replace $\e^kt$ by $t_k$ again.

\begin{verbatim}
PerturbationEqns[linearpart_,nonlinearpart_,fun_Symbol,
    vars_List,time_Symbol,ord_Integer] :=
  Module[{i,j,e,eq},
    eq = ((linearpart == -e*nonlinearpart)
             /. {fun->Sum[e^i*fun[i][time,Sequence@@Table[e^j*
                                       time,{j,ord}],Sequence@@
                                       DeleteCases[vars,time]],
                          {i,0,ord}]});
    eq = (eq /. ((Dt[#, __]->0)& /@ Join[vars,{time,e}]));
    eq = (Equal@@#)& /@
             Transpose[Take[CoefficientList[#,e],1+ord]& /@
                         (List@@eq)];
    eq /. Table[e^j*time->time[j],{j,ord}]
    ]
\end{verbatim}

\subsubsection{Perturbation Equations, Given Linear Mode}

In step two we set $\psi_0(t,t_1,\vec{x})=\sum_{j=1}^s
A_j(t_1)\varphi_j(\vec{x},t)$ as described above. This is done by
the function

{\tt PlugInGenericWaveTuple[eq, $\psi$, \{x$_1$,...,x$_n$\}, t, A,
B, $\theta$, s]}

where the first argument is the output of the previous step. The
symbols {\tt B} and $\theta$ have to be passed for labeling the
shape and phase functions respectively. The output consists of two
parts. The first part of the list formulates the assumption
$L(\varphi_j)=0$ explicitly for each of the waves. This is not used
in subsequent computations, but is provided as a way to check the
assumption. The second part of the list is the equation
corresponding to the coefficients of $\e$ from the previous step,
with $\psi_0$ as above. As the task of this step is so short the
implementation does not need further explanation.
\begin{verbatim}
PlugInGenericWaveTuple[eq_List,fun_Symbol,vars_List,
  t_Symbol,A_Symbol,B_Symbol,theta_Symbol,num_Integer] :=
  Module[{i,j,waves,n=Length[DeleteCases[vars,t]]},
    waves = Table[A[j][Slot[2]]*
                    Product[B[i][j][Slot[i+2]],{i,n}]*
                    Exp[I*theta[j][Sequence@@Table[Slot[i+2],
                                       {i,n}],Slot[1]]],
                  {j,num}];
    {Table[eq[[1]] /. fun[0]->Function[Evaluate[waves[[j]]]],
           {j,num}],
     Expand /@
         (eq[[2]] /. fun[0]->Function[Evaluate[Total[waves]]])
    }]
\end{verbatim}

\subsubsection{Time and Scale Averaging}

Step three is the most elaborate. Under the assumption that
interchange of averaging over time and inner product is justified,
an integrand
\[h=\lim_{T\to\infty}\frac{1}{T}\int_0^T\psi_0\overline{\varphi_k}\,dt\]
is computed that when integrated over the domain yields
\[\int\limits_\Omega h=\lim_{T\to\infty}\frac{1}{T}\int_0^T
\langle\psi_0,\varphi_k\rangle_{L^2(\Omega)}\,dt.\] Resonance
conditions posed on the phase functions are explicitly used by
\begin{quote}
\ttfamily
Resonance[eq, linwav, \{x$_1$,..,x$_n$\}, t, \\
\strut~~~
\{$\lambda_1$,..,$\lambda_p$\}, A, B, $\theta$, s, cond, k]
\end{quote}
which receives the output from the previous step in {\tt eq}. Here
{\tt cond} specifies the resonance condition in terms of the
$\theta_j$, which have to be entered as {\tt
$\theta$[j][x$_1$,..,x$_n$,t]} respectively. The last argument is
the index of the wave $\varphi_k$ in the integral above.
Alternatively {\tt Resonance2} uses explicit parameter settings {\tt
paramvalues} for the waves instead of {\tt cond}. This has been
necessary because the general {\tt Resonance} does not give useable
results (see Section \ref{obst} for more details). The main work in
this step is to find out which terms do not contribute to the
result. We exploit the fact that oscillating terms vanish when
averaged over time by simply omitting those summands of
$\langle\psi_0,\varphi_k\rangle_{L^2(\Omega)}$ that have a factor
$\exp(i\theta)$ with some time-dependent phase $\theta$. The code
for {\tt Resonance} is not shown here, but is quite similar to {\tt
Resonance2}.

\begin{verbatim}
Resonance2[eq_List,linwav_List,vars_List,t_Symbol,params_List,
    A_Symbol,B_Symbol,theta_Symbol,num_Integer,
    paramvalues_List,testwave_Integer] :=
  Module[{e,i,j,n=Length[DeleteCases[vars,t]]},
    e = Expand[(List@@Last[eq])*
               Exp[-I*theta[testwave][Sequence@@
                                          DeleteCases[vars,t],
                                      t]]];
    e = e /.
         Table[
           theta[j] ->
             (Evaluate[(linwav[[n+1]] /.
                         (Rule@@#& /@
                           Transpose[{params,paramvalues[[j]]}]
                         )
                       ) /. Append[Table[
                                     DeleteCases[vars,t][[i]]
                                       -> Slot[i],
                                     {i,n}],
                                   t -> Slot[n+1]]
                      ]&
             ),
           {j,num}];
    e = MapAt[
          (Function[theta,If[FreeQ[theta,t],theta,0]
                   ]
               [Simplify[#]]
          )&,
          e,
          Position[e,Exp[_]]];
    e = Equal@@
          (e*Conjugate[A[testwave]][t[1]]*
            Product[Conjugate[B[i]
                               [testwave]
                               [DeleteCases[vars,t][[i]]]
                             ],
                    {i,n}]
          ) /.
            Flatten[
              Table[B[i][j] ->
                      Function[
                        Evaluate[DeleteCases[vars,t][[i]]],
                        Evaluate[linwav[[i]] /.
                                   (Rule@@#& /@
                                     Transpose[
                                       {params,paramvalues[[j]]
                                       }]
                                   )]],
                    {i,n},{j,num}]]
    ]
\end{verbatim}

The integration of $h$ is done by Mathematica and can be quite
time-consuming. So {\tt ODESystem} simplifies the integrand first to
make integration faster. Still the expressions involved can be quite
complicated. This is the most time-consuming part during
construction of the ODEs.

\subsection{Obstacles}\label{obst}

Mathematica sometimes does not seem to take care of special cases
and consequently has problems with evaluating expressions depending
on symbolic parameters. We give two simple examples to illustrate
this issue:
\begin{itemize}
\item{} Orthogonality of sine-functions.\\
  Indeed, it holds that $$\forall m,n\in\mathbb{N}: \int_0^{2\pi}\sin(mx)\sin(nx)dx=\pi\delta_{m,n}.$$
  Computing this in Mathematica by
\begin{quote}
\ttfamily
 Integrate[Sin[m*x]Sin[n*x], \{x,0,2$\pi$\}, \\
\strut~~~
  Assumptions $\to$ m$\in$Integers \&\& n$\in$Integers]
\end{quote}
 yields 0 independently of $m,n$ instead.

\item{} Computation of a limit.\\
Mathematica evaluates an expression
 $$\forall n\in\mathbb{Z}:\lim_{x{\to}n}\frac{\sin(x\pi)}{x}=\pi\delta_{n,0}$$
 and similar expressions in two different ways getting two different answers.
 On the one hand

 {\tt Limit[Sin[(m-n)$\pi$]/(m-n), m$\to$n,

 Assumptions $\to$ m$\in$Integers \&\& n$\in$Integers]}

 gives 0. On the other hand, however, when the condition $m,n\in\mathbb{Z}$ is not
 used for computing the result Mathematica yields the correct answer
 $\pi$, as with

   {\tt Limit[Sin[(m-n)$\pi$]/(m-n), m$\to$n]}.

\end{itemize}
Unfortunately these issues prevented us from obtaining a nice formula for the
coefficients in symbolic form by {\tt Resonance}. So we just compute results
for explicit parameter settings using {\tt Resonance2}.

\subsection{Results}

\subsubsection{Atmospheric Planetary Waves}

For the validation of our program we consider the barotropic
vorticity equation on the sphere first. Here numerical values of the
coefficients $ \a_i $ are available (Table 1, \cite{KL-06}). The
equation looks quite similar
$$\frac{\partial \triangle \psi}{\partial t} + 2\frac{\partial \psi}{\partial\lambda}= -\e J(\psi, \triangle \psi)$$
However in spherical coordinates
($\phi\in[-\frac{\pi}{2},\frac{\pi}{2}]$, $\lambda\in[0,2\pi]$) the
differential operators are different:
$$\triangle = \frac{\partial^2}{\partial\phi^2}+\frac{1}{\cos(\phi)^2}
\frac{\partial^2}{\partial\lambda^2}-\tan(\phi)\frac{\partial}{\partial\phi}$$
$$J(a,b)=\frac{1}{\cos(\phi)}\left(\frac{\partial a}{\partial\lambda}
\frac{\partial b}{\partial\phi}-\frac{\partial
a}{\partial\phi}\frac{\partial b}{\partial \lambda}\right).$$ The
linear modes  have in this case  the following form \cite{ped} \be
\label{spherical_mode}P_n^m(\sin(\phi))\exp(i(m\lambda+\frac{2m}{n(n+1)}t))\ee
where $P_n^m(\mu)$ are the associated Legendre polynomials of degree
$n$ and order $m\le n$, so again they depend on the two parameters
$m$ and $n$. Also resonance conditions on the parameters look
different in this case.

Now we compute the coefficient $\alpha_3$ in (\ref{dynsys}).
In \cite{KL-06} we find the following equation for the amplitude $A_3$
$$n_3(n_3+1)\frac{\partial A_3}{\partial t_1}(t_1)=2iZ(n_2(n_2+1)-n_1(n_1+1))A_1(t_1)A_2(t_1)$$
so $\alpha_3=2iZ\frac{n_2(n_2+1)-n_1(n_1+1)}{n_3(n_3+1)}$. Parameter
settings and corresponding numerical values for $Z$ were taken from
the table below (see \cite{KL-06}). For this equation and $s=3$
results produced by our program have the form
$c_1\overline{A_3}\dot{A_3}=c_2A_1A_2\overline{A_3}$, so
$\alpha_3=c_2/c_1$.

Testing all resonant triads from the Table 1 from \cite{KL-06}, we
see that the coefficients differ merely by a constant factor of
$\pm\sqrt{8}$ which is due to the different scaling of the Legendre
polynomials. In our computation they were normalized s.t.
$\int_{-1}^1P_n^m(\mu)^2d\mu=1$. With three triads, however, results
were completely different. Interestingly this were exactly those
triads for which no $\varphi_0$ appears in the table.

Furthermore, for the other coefficients in (\ref{dynsys}) our
program computes $\alpha_1=\alpha_2=0$ in all tested parameter
settings. This fact can be easily understood in the following way.
We checked only resonance conditions but not the conditions for the
interaction
coefficients to be non-zero which are elaborated enough:%
 $$
 m_i \leq
n_i,  \quad  n_i \neq n_j \quad \forall i=1,2,3, \quad |n_1-n_2| <
n_3 < n_1+n_2,$$ and $$ n_1+n_2+n_3 \quad \mbox{is odd}.$$ Randomly
taken parameter setting does not satisfy these conditions.

\subsubsection{Ocean Planetary Waves}

Returning to the original example on the domain
$[0,L_x]\times[0,L_y]$, we find explicit formulae for the
coefficients in \cite{rectangular}. According to Section \ref{obst}
we can only verify special instances and not general formulae.

Linear modes  have now the form \cite{rectangular} \be
\label{rectangular_mode}\sin(\pi\frac{mx}{L_x})\sin(\pi\frac{ny}{L_y})\exp(i(\frac{\b}{2\o}x+\omega
t))\ee with $m,n\in\mathbb{N}$ and
$\omega=\frac{\beta}{2\pi\sqrt{(\frac{m}{L_x})^2+(\frac{n}{L_y})^2}}$.

Parameter settings solving the resonance conditions were computed as in section \ref{sec:WaveNumbers}.
Unfortunately results do not match and we have no explanation for that. In particular the condition
 $\frac{\alpha_1}{\o_1^2}+\frac{\alpha_2}{\o_2^2}+\frac{\alpha_3}{\o_3^2}=0$ stated in \cite{rectangular}
 does not hold for the results of our program since we got $\alpha_1=\alpha_2=0$ in all tested parameter settings,
 just as in the spherical case.

For example, if we try the triad {\tt \{\{2,4\},\{4,2\},\{1,2\}\}} where $L_x=L_y=1$ our program computes
$\alpha_3=\frac{32\sqrt{5}}{11}\pi\left(\sin(3\sqrt{5}\pi)-i(1+\cos(3\sqrt{5}\pi))\right)$,
whereas the general formula yields $\alpha_3=\frac{19+7\sqrt{5}}{11}\pi\sin(3\sqrt{5}\pi)$.
However, if we use a triad with $q=1$, e.g. {\tt \{\{24,18\},\{9,12\},\{8,6\}\}}, both agree on $\alpha_1=\alpha_2=\alpha_3=0$.

\section{Resonance Conditions}
\label{sec:WaveNumbers}

The main equation to solve is
$$
\frac{1}{\sqrt{(\frac{m_1}{L_x})^2+(\frac{n_1}{L_y})^2}}+\frac{1}{\sqrt{(\frac{m_2}{L_x})^2+(\frac{n_2}{L_y})^2}}=
\frac{1}{\sqrt{(\frac{m_3}{L_x})^2+(\frac{n_3}{L_y})^2}}
$$
for all possible $\ m_i, n_i \in \Integer \ $ with the scales $L_x$
and $L_y$ (also $\in \Integer \ $) and then to check the  condition
$n_1\pm n_2=n_3$. In the following argumentation it will be seen
that $L_x$ and $L_y$ can be assumed to be free of common factors.
Below we refer to $L_x$ and $L_y$ as to {\it the scale
coefficients}.

The first step of the algorithm  implemented in Mathematica is to
rewrite the equation to $
\frac{1}{\sqrt{\tilde{m_1}^2+\tilde{n_1}^2}}+\frac{1}{\sqrt{\tilde{m_2}^2+\tilde{n_2}^2}}=
\frac{1}{\sqrt{\tilde{m_3}^2+\tilde{n_3}^2}} $ and transform it in
the following way: we factorize the result of each
$\tilde{m_i}^2+\tilde{n_i}^2$ and obtain  with
$\rho_1\cdot\ldots\cdot\rho_r$ being the factors of $m_i^2+n_i^2$
and $\a_1\cdot\ldots\cdot\a_r$ their respective powers:
$$m_i^2+n_i^2=\rho_1^{\a_1}\cdot\rho_2^{\a_2}\cdot\ldots\cdot\rho_r^{\a_r}.$$
We will now define {\it a weight $\ \g_i\ $ of the wave-vector } $\ (m_i, n_i) \ $ as the product of the $\rho_j$'s to the quotient
of their respective $\a_j$ and 2. The weight $q_i$ will be the name of the product of
the $\rho_j$'s which have an odd exponent:
$$
\sqrt{m_i^2+n_i^2}=\g_i\sqrt{q_i}.
$$
Our equation then can be re-written as
$$
\frac{1}{\g_1\sqrt{q_1}}+\frac{1}{\g_2\sqrt{q_2}}=\frac{1}{\g_3\sqrt{q_3}}
$$
and one easily sees that the only way for the equation to possibly
hold is $\ q_1=q_2=q_3=q \ $ (see \cite{comp-all1} for details).
Further we call $ q $ {\it an index} of the corresponding
wave-vectors. The set of all wave-vectors with the same index is
called {\it a class of index $\ q\ $}  and is denoted as $\ Cl_q. \
$ Obviously, the solutions of the resonance conditions are to be
searched for with separate classes only.

 At this point one can also see that only
such scales, $L_x$ and $L_y$, without common factors are reasonable. If they
had a common factor, it would cancel out in the equation.

\subsection{Method Description}
The following five steps are the main steps of the algorithm:
\begin{itemize}
\item{} {\bf Step 1:} Compute the list of all possible indexes $q$.

To compute the list of all indexes $\ q,\ $ we  use the fact that
they have to be square-free and each factor of $\ q\ $ has to be
different from $3\mod 4$ (Lagrange theorem). There exist  57
possible possible indexes in our computational domains $\ q\le 300:\
$
\begin{eqnarray}
&&\{1, 2, 5, 10, 13, 17, 26, 29, 34, 37, 41, 53, 58, 61, 65, 73, 74,
82,
85,89, \nonumber \\
&&97,101,106,109, 113, 122, 130, 137, 145, 146, 149, 157, 170, 173,
178,
\nonumber \\
&&181,185, 193, 194, 197,202, 205, 218, 221, 226, 229, 233, 241, 257,
\nonumber\\
&& 265, 269, 274, 277, 281, 290, 293, 298\nonumber\}
\end{eqnarray}

\item{} {\bf Step 2:} Solve the weight equation
$\frac{1}{\g_1}+\frac{1}{\g_2}=\frac{1}{\g_3}$.

For solving the weight equation, we transform it into the equivalent
form:
\begin{equation}\g_3=\frac{\g_1\:\g_2}{\g_1+\g_2}\label{gamma3}\end{equation}
The solution triples $\{\g_1,\g_2,\g_3\}$ can now be found  by the
two for-loops  over $\g_1$ and $\g_2$ up to a certain maximum
parameter and $\g_3$ is then being founded constructively with
formula (\ref{gamma3}).

\item{} {\bf Step 3:} Compute all possible pairs $(m_i,n_i)$ - if there are any - that
satisfy $m_i^2+n_i^2=\g_i^2\: q$.

To compute our initial variables $\ m_i, n_i, \ $ we use the
Mathematica standard function \textbf{SumOfSquareRepresentation[d,
x]} which produces  a list of all possible representations of an
integer $x$ as a sum of $d$ squares, i.e. we can find all possible
pairs $\ (a,b)\ $ with $\ d=2\ $ such that they satisfy $\
a^2+b^2=x.\ $
 Therefore,
checking the  condition $m_i^2+n_i^2=\g_i^2\: q$ is easy.

\item{} {\bf Step 4:} Sort out the solutions $\{m1,n1,m2,n2,m3,n3\}$ that do not fulfill the
condition $n1\pm n2=n3$.

\item{} {\bf Step 5:} Check if by dividing the $m_i$ by $L_x$ and the $n_i$ by $L_y$
there are still exist some solutions.

Last two steps are trivial.
\end{itemize}

\subsection{The Implementation}
 Our implementation is quite straightforward and the main program is based on 4 auxiliary
functions shown in the following subsections.

\subsubsection{List of Indexes}
The function \textbf{constructqs[max]} produces the list of all
possible indexes $\ q\ $ up to the parameter $max$. The first
(obvious) $q$'s $sol=\{1\}$ is given and the function checks the
conditions starting with $n=2$. Every time $n$ satisfies the
conditions, it is appended to the list $sol$. If one condition
fails, the next $n=n+1$ is considered and so on until $n$ reaches
the  parameter $max$. Then the list $sol$ is returned:

{\tt Clear[constructqs];

constructqs[n\_, sol\_List, max\_]; n>max := sol (*6*)\\
constructqs[n\_?SquareFreeQ, sol\_List, max\_] \\ :=
  constructqs[n+1, Append[sol, n], max] (*5*)

constructqs[n\_?SquareFreeQ, sol\_List, max\_]; \\
  MemberQ[Mod[PrimeFactorList[n], 4], 3]\\
   := constructqs[n+1, sol, max] (*4*)

constructqs[n\_, sol\_List, max\_]; !SquareFreeQ[n]\\
 := constructqs[n+1, sol, max] (*3*)

constructqs[1] := \{1\}  (*2*)

 constructqs[max\_] :=
constructqs[3, \{1\}, max] (*1*)}

\subsubsection{Weight Equation}
 The function \textbf{find$\g$s[$\g$max]}
solves the weight equation in the following way. For a fixed
 $ \g_1 $ and $ \g_2 $ running between 1
and $\g max$, it is checked if $\ \g_3\ $  is an integer. If it is,
the triple $\ \{\g_1,\g_2,\g_3\}\ $ is added to the list $\ sol\ $
which is empty at the initial moment. Once $\ \g_2\ $ reaches $\ \g
max,\ $ it is set to 1 again and the search starts again with
$\g_1=\g_1+1$. This is done as long as both $\ \g_1\ $ and $\ \g_2\
$ are lower than $\ max.\ $ Finally the list $\ sol\ $ is
returned:

{\tt find$\g$s[$\g$max\_, $\g$1\_, $\g$2\_, sol\_List];

 $\g$1 >
$\g$max :=
(Clear[$\g$3],sol) (*6*)

find$\g$s[$\g$max\_, $\g$1\_, $\g$2\_, sol\_List]; ($\g1\leq \g$max
\&\&
$\g$2>$\g$max \&\& \\
IntegerQ[$\g$3=($\g$1$\g$2)/($\g$1+$\g$2)])\\
 := find$\g$s[$\g$max,
$\g$1+1, 1, Append[sol, \{$\g$1, $\g$2, $\g$3\}]] (*5*)

find$\g$s[$\g$max\_, $\g$1\_, $\g$2\_, sol\_List];\\
 ($\g1\leq\g$max
\&\& $\g$2>$\g$max \&\&
\\!IntegerQ[$\g$3=($\g$1$\g$2)/($\g$1+$\g$2)]) \\:=
find$\g$s[$\g$max, $\g$1 + 1, 1, sol] (*4*)

find$\g$s[$\g$max\_, $\g$1\_, $\g$2\_, sol\_List];\\
($\g1\leq\g$max
\&\& $\g2\leq\g$max \&\& IntegerQ[$\g$3=($\g$1$\g$2)/($\g$1+$\g$2)])
\\:= find$\g$s[$\g$max, $\g$1, $\g$2 + 1, Append[sol, \{$\g$1,
$\g$2, $\g$3\}]]
(*3*)

find$\g$s[$\g$max\_, $\g$1\_, $\g$2\_, sol\_List]; \\
($\g1\leq\g$max
\&\& $\g2\leq \g$max \&\&
!IntegerQ[$\g$3=($\g$1$\g$2)/($\g$1+$\g$2)]) \\:=
find$\g$s[$\g$max, $\g$1, $\g$2 + 1, sol] (*2*)

find$\g$s[$\g$max\_] := find$\g$s[$\g$max, 1, 1, \{\}]) (*1*)
}

For \textbf{find$\g$s[$\g$max]} to be executable, the iteration depth of
$2^{12}$ is not sufficient and it was set to $\ \infty.$

\subsubsection{Linear Condition}
The third auxiliary function \textbf{makemns} checks whether the linear condition
$\ n_1\pm n_2=n_3\ $ is fulfilled and
 structures the solution set into a list of pairs
$\{\{m_1,n_1\},\{m_2,n_2\},\{m_3,n_3\}\}: $

{\tt
Clear[makemns];

makemns[m1\_, n1\_, m2\_, n2\_, m3\_, n3\_] := \{\} (*3*)

makemns[m1\_, n1\_, m2\_, n2\_, m3\_, n3\_];\\
 (n1 + n2 == n3 $\|$
n1 - n2 == n3) := \\
\strut~~~
\{\{m1, n1\}, \{m2, n2\}, \{m3, n3\}\} (*2*)

makemns[mn1\_List, mn2\_List, mn3\_List] := \\
\strut~~~ Cases[Flatten[Table[makemns[mn1[[i,1]], mn1[[i,2]], \\
\strut~~~~~~ mn2[[j,1]],
mn2[[j,2]], mn3[[k,1]], mn3[[k,2]]],\\
\strut~~~~~~ \{i, 1, Length[mn1]\}, \{j,
1, Length[mn2]\}, \\
\strut~~~~~~ \{k, 1, Length[mn3]\}], 2], \\
\strut~~~~~~ \{\{x1\_,x2\_\},
\{x3\_,x4\_\}, \{x5\_,x6\_\}\}] (*1*) }

The function \textbf{makemns} is called three times:

In (*1*) from 3 lists  of arbitrarily many pairs \{mi, ni\}, a
3-dimensional array is made combining entries of the 3 lists with
each other. Each entry calls the same program with the parameters of
the current combination of \{m1,n1,m2,n2,m3,n3\}.

In (*2*) and (*3*) it is decided whether the condition $\ n1\pm
n2=n3\ $ is fulfilled.  If it is, a solution
\{\{m1,n1\},\{m2,n2\},\{m3,n3\}\} is written in the array.  The
table is then flattened to the level 2 in order to have a list of
solutions. In the end, all empty lists have to be sorted out, done
by the function \textbf{Cases} which keeps only those cases that
have the shape \{\{x1\_,x2\_\},\{x3\_,x4\_\},\{x5\_,x6\_\}\}.

\subsubsection{Scale Coefficients}
Finally, the function \textbf{respectL[sol, Lx, Ly]} divides each component
of the solution by the pair $(L_x,L_y)$ and sorts out the result if any of
the 6 components does not remain an integer:

{\tt respectL[sol\_List, Lx\_, Ly\_] :=\\
\strut~~~ Map[solution[\#]\&, \\
\strut~~~~~~
    Cases[Map[\#/\{Lx, Ly\}\&,\\
\strut~~~~~~
        Map[\#[[1]]]\&,
          sol], \{2\}], \{\{\_Integer, \_Integer\}, \\
\strut~~~~~~\{\_Integer,
\_Integer\}, \{\_Integer, \ \_Integer\}\}]] }

The function \textbf{respectL[sol, Lx, Ly]} gets as an input the list of the form
\{solution[\{\{m1,n1\},\{m2,n2\},\{m3,n3\}\}],...\} and returns the list of the same form.

\subsection{Results}
\label{results}

All solutions in the computation domain $\ m,n \le 300\ $ have been
found in a few minutes. Notice that computations in the domain $\
m,n \le 20\ $ by direct search, without introducing indexes $\ q \ $
and classes $\ Cl_q\ $ took about 30 minutes. A direct search in the
domain $\ m,n \le 30\ $ has  been interrupted after 2 hours, since
no results were produced.

The number of solutions depends drastically on the scales $\ L_x \ $
and $\ L_y,\ $ some data are given below (for the domain $\ m,n \le
50:\ $)

$(L_x=1, L_y=1):$  76 solutions;

$(L_x=3, L_y=1):$  23 solutions;

$(L_x=6, L_y=16):$  2 solutions;

$(L_x=5, L_y=21):$  2 solutions;

$(L_x=11, L_y=29):$  no solutions (search up to 300, for both $qmax$ and $\g max$).

Interestingly enough, in all tried possibilities, only an odd $q$
yield solutions.

\section{Structure of the Solution Set}

\subsection{Method Description}

The graphical way to present 2D-wave resonances suggested in
\cite{AMS} for 3-wave interactions is to regard each 2D-vector $\
\vec{k}=(m,n)\ $ as a node $\ (m,n)\ $ of integer lattice in the
spectral space and connect those nodes which construct one solution
(triad, quartet, etc.). Having computed already all the solutions of
(\ref{res}) in Section 4, now we are interested in the structure of
resonances in spectral space. To each node $\ (m,n)\ $ we can
prescribe an amplitude $\ A(m,n, t_1)\ $ whose time evolution can be
computed from the dynamical equations obtained in Section 3. Thus,
 solution set of resonance conditions (\ref{res}) can be thought of as a collection of triangles, some of
 them are isolated, some form small groups connected by one or two vertices. Corresponding dynamical systems
 can be re-constructed from the structure of these groups.
For instance, a single isolated triangle corresponding to a solution
with wave vectors  $(m_1,n_1)(m_2,n_2)(m_3,n_3)$ and wave amplitudes
$\{(A1,A2,A3)\}$ corresponds to the following dynamical system:
\begin{eqnarray*}
  \dot{A_1} & = & \alpha_1 A_2 A_3 \\
  \dot{A_2} & = & \alpha_2 A_1 A_3 \\
  \dot{A_3} & = & \alpha_3 A_1 A_2
\end{eqnarray*}
with $\alpha_i$ being functions of all $m_i, n_i$ (see Section 3).

If that two triangles share one common vertex
$\{(A1,A2,A3),(A3,A4,A5)\},$ the the corresponding dynamical system
is
\begin{eqnarray*}
  \dot{A_1} & = & \alpha_1 A_2 A_3 \\
  \dot{A_2} & = & \alpha_2 A_1 A_3 \\
  \dot{A_3} & = &   \alpha_{3,1} A_1 A_2 + \alpha_{3,2} A_4 A_5  \\
  \dot{A_4} & = & \alpha_4 A_3 A_5 \\
  \dot{A_5} & = & \alpha_5 A_3 A_4
\end{eqnarray*}
If two triangles have two vertices  in common
$\{(A1,A2,A3),(A2,A3,A4)\}$, then the dynamical system is quite
different:
\begin{eqnarray*}
  \dot{A_1} & = & \alpha_1 A_2 A_3 \\
  \dot{A_2} & = &   \alpha_{2,1} A_1 A_3 + \alpha_{2,2} A_3 A_4  \\
  \dot{A_3} & = &   \alpha_{3,1} A_1 A_2 + \alpha_{3,2} A_2 A_4  \\
  \dot{A_4} & = & \alpha_4 A_2 A_3 = \frac{\alpha_4}{\alpha_1} \dot{A_1}
\end{eqnarray*}
Using the fourth equation, the formulae for $\dot{A_2}$ and $\dot{A_3}$ can be simplified to:
\begin{eqnarray*}
  \dot{A_4} & = & \frac{\alpha_4}{\alpha_1} \dot{A_1}
\Rightarrow A_4  =  \frac{\alpha_4}{\alpha_1} A_1 + \beta_1  \\
  \dot{A_2} & = &   A_1 A_3 \left( \alpha_{2,1} + \frac{\alpha_{2,2} \alpha_4}{\alpha_1} \right)                  + \frac{\alpha_4 \beta_1}{\alpha_1}  \\
  \dot{A_3} & = &  A_1 A_2 \left( \alpha_{3,1} + \frac{\alpha_{3,2} \alpha_4}{\alpha_1} \right)                  + \frac{\alpha_4 \beta_1}{\alpha_1}
\end{eqnarray*}

This means that {\it qualitative dynamics} of the 3-term mesoscopic system depends
{\it not on the geometrical structure} of the solution set but on its {\it topological structure}.
 Constructing the topological structure of the solution set, we do not consider concrete values of
 the solution but only the way how triangles are connected. In any
 finite spectral domain we can compute all independent wave clusters
 and write out corresponding dynamical systems thus obtaining
 complete information about energy transfer through the spectrum.
Of course, {\it quantitative}
 properties of the dynamical systems
 depend on the specific values of $\ m_i, n_i\ $ (for instance, values of interaction coefficients $\ \a_i,\ $
 magnitudes of periods of the energy exchange among the waves belonging to one cluster, etc.)

\subsection{Implementation}
\label{sec:TopologicalImplementation}

To construct the topological structure of a given solution set we
need first to find all groups of connected triangles. This is done
by the following procedure:

\begin{verbatim}
FindConnectedGroups[triangles_List] :=
 Block[{groups = {}, tr = triangles, newgroup},
   While[Length[tr] > 0,
     {newgroup, tr} =
       FindConnectedTriangles[{First[tr]}, Rest[tr]];
     groups = Append[groups, newgroup];
   ];
   groups
 ];

FindConnectedTriangles[grp_List,triangles_List]:=
 Module[{points,newGrpMember,tr=triangles},
   points=Flatten[Apply[List,grp,2],1];
   newGrpMember=Cases[tr, _[___,#1,___]]&/@points;
   (tr=DeleteCases[tr, _[___,#1,___]])&/@points;
   newGrpMember=Union[Join@@newGrpMember];
   If[Length[newGrpMember]==0,
     {grp,tr},
     newGrpMember=FindConnectedTriangles[newGrpMember,tr];
     {Join[grp,First[newGrpMember]],
      newGrpMember[[2]]}
    ]
 ];
\end{verbatim}

The function \verb|FindConnectedGroups| expects a list of triangles
as input, and three  different types for data structure can be used.
The first type is just a list of three pairs, where each pair
contains the coordinates of a node, for example
\verb|{{1,2},{3,4},{5,6}}|. An alternative type is like the type
before just with another
head symbol instead of list, e.g.\\ \verb|Triangle[{1,2},{3,4},{5,6}]|.

 The function also works for vertex numbers instead of coordinates, e.g. \verb|Triangle[1, 2, 3]|.
 In every case the function returns a partition of the input list where all elements of a list are
 connected and elements of different lists have no connection to each
 other.

The function \verb|FindConnectedTriangles| is an auxiliary function
which has two parameters. The first list contains allconnected
triangles. The second list contains all other triangles which are
possibly connected to one of the triangles in the first list. The
function \verb|FindConnectedTriangles| returns a pair of lists: the
first list contains all triangles which are  connected to the
selected triangles, the second list contains all remaining.

The input list for \verb|FindConnectedTriangles| is a list of
3-element lists. Before we can use the results produced in Section
\ref{sec:WaveNumbers} as an input we have to transform the data.
This can be easily done by:
\begin{verbatim}
TransformSolution[sol_List]:=
  Flatten[Rest/@sol]/.solution[trs:{___List}]->trs;
\end{verbatim}

\paragraph*{Some remarks on the implementation.}\mbox{}

The function \verb|FindConnectedGroups| selects a triangle,
which is not yet in a group and calls the function \verb|FindConnectedTriangles|.
Since the returned first list always contains at least one triangle, the length of
the list \verb|tr| decreases in every loop call, hence the \verb|FindConnectedGroups| terminates.
The question left is how to find all triangles connected with a certain triangle.
This has been done in the following way. First we search for all triangles which share
at least one node with this triangle. Then we restart the search with all triangles found.
For efficiency reasons it is better to perform the search with all triangles we found in one
step together. If in one step no further triangles are found then we are ready and
return the list of connected triangles and the remaining list. In each step we
remove all triangles we found from the list of triangles which are not declared
as connected. This increases the speed because the search is faster if there are less elements to compare.
More important, this prevent us to search in loops and find some triangles more than once. In general, search
in a loop can be the reason for a termination problem but due to shrinking the list of triangles to search
for in every step the termination can be guaranteed.

\subsection{Results}

\begin{figure}[htb]
\begin{center}
\includegraphics[width=10cm]{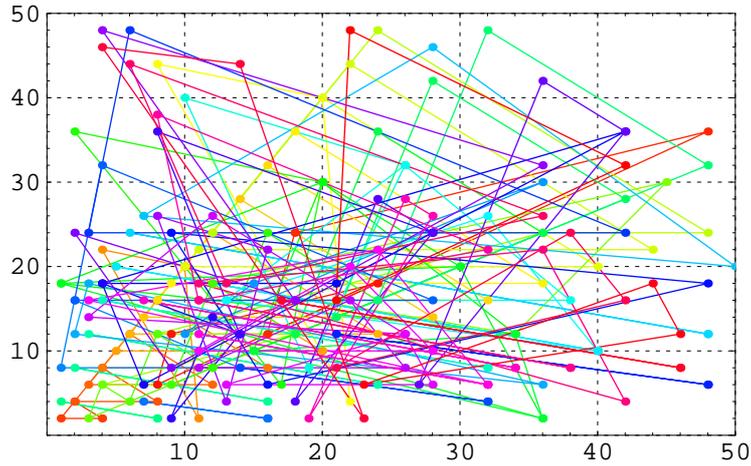}
\caption{The geometrical structure of the result in domain $D = 50$}
\label{fig:ResultGeometrical}
\end{center}
\end{figure}

In the Figure \ref{fig:ResultGeometrical} the geometrical structure
of the solution set is shown, for the case $\ m_i, n_i\ \le 50\  $
and $\ L_x = L_y = 1.$

Below we show all the topological elements of this solution set.

{\bf 1.} 21 groups contain only one triangle (obviously, they have isomorphic dynamical systems):
\[
\begin{array}{ll}
 \{\{3,18\},\{36,6\},\{2,12\}\} &
 \{\{4,46\},\{14,44\},\{23,2\}\}  \\
 \{\{6,44\},\{36,26\},\{13,18\}\} &
 \{\{6,48\},\{42,24\},\{3,24\}\} \\
 \{\{8,26\},\{16,22\},\{13,4\}\} &
 \{\{9,24\},\{48,18\},\{16,6\}\} \\
 \{\{14,28\},\{28,14\},\{7,14\}\} &
 \{\{18,36\},\{36,18\},\{9,18\}\} \\
 \{\{22,16\},\{26,8\},\{11,8\}\}  &
 \{\{22,20\},\{28,10\},\{11,10\}\} \\
 \{\{22,44\},\{44,22\},\{11,22\}\} &
 \{\{22,48\},\{42,32\},\{21,16\}\} \\
 \{\{24,18\},\{9,12\},\{8,6\}\}   &
 \{\{26,28\},\{28,26\},\{19,2\}\} \\
 \{\{28,42\},\{42,28\},\{21,14\}\} &
 \{\{28,46\},\{50,20\},\{7,26\}\} \\
 \{\{36,22\},\{42,4\},\{11,18\}\} &
 \{\{36,30\},\{15,18\},\{10,12\}\} \\
 \{\{38,24\},\{42,16\},\{21,8\}\} &
 \{\{44,18\},\{46,12\},\{23,6\}\} \\
 \{\{48,36\},\{18,24\},\{16,12\}\}
\end{array}
\]
\begin{center}
\includegraphics[width=4cm]{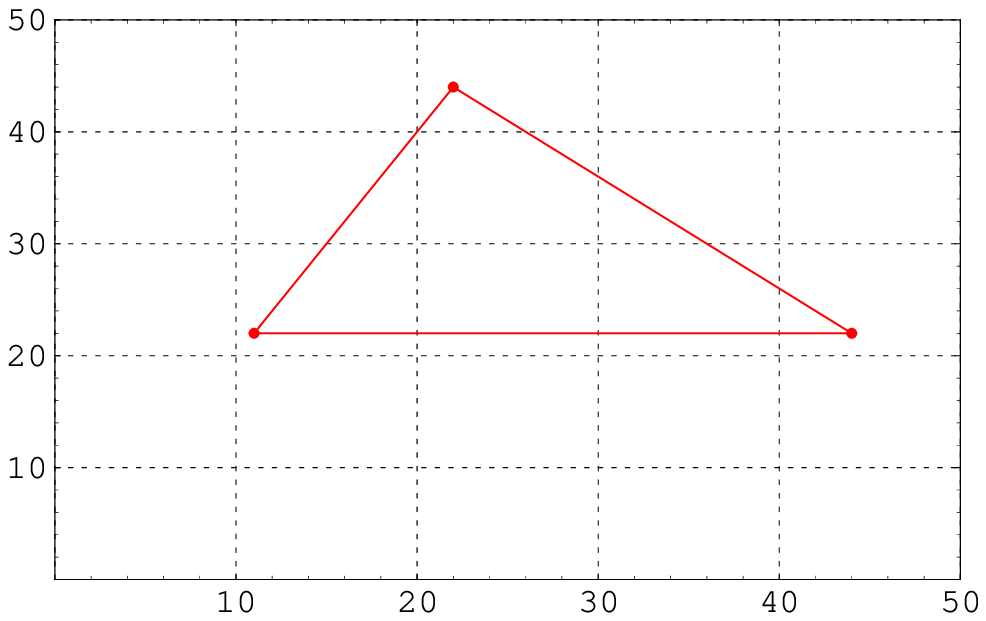}
\end{center}

{\bf 2.} Further 9 groups contain also one triangle, but in each triangle two points coincide
(again, they have isomorphic dynamical systems):
\[
\begin{array}{ll}
 \{\{8,2\},\{8,2\},\{1,4\}\} &
 \{\{16,2\},\{16,2\},\{7,4\}\} \\
 \{\{16,4\},\{16,4\},\{2,8\}\} &
 \{\{24,6\},\{24,6\},\{3,12\}\} \\
 \{\{32,8\},\{32,8\},\{4,16\}\} &
 \{\{34,8\},\{34,8\},\{7,16\}\} \\
 \{\{46,8\},\{46,8\},\{17,16\}\} &
 \{\{48,6\},\{48,6\},\{21,12\}\} \\
 \{\{48,12\},\{48,12\},\{6,24\}\}
\end{array}
\]
\begin{center}
\includegraphics[width=4cm]{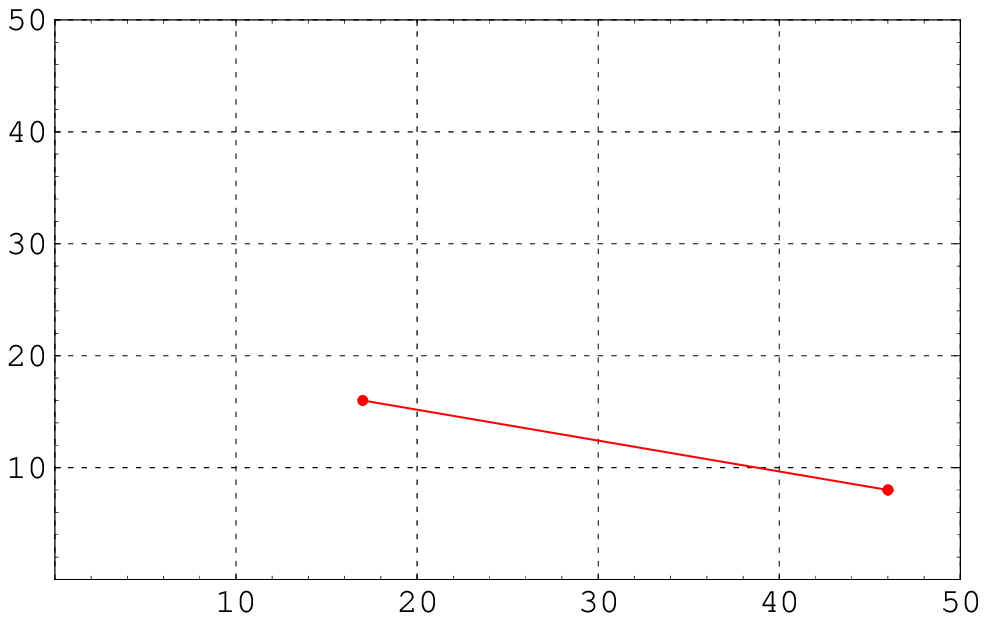}
\end{center}

{\bf 3.} There exist 2 groups with two triangles each (by observation of the geometrical pictures
it is easy to determine that both have isomorphic dynamical systems):
\[
\begin{array}{llll}
 \{& \{\{2,24\},\{18,16\},\{9,8\}\},   & \{\{4,48\},\{36,32\},\{18,16\}\} & \} \\
 \{& \{\{12,26\},\{26,12\},\{3,14\}\}, & \{\{26,12\},\{28,6\},\{13,6\}\}  & \}
\end{array}
\]
\begin{center}
\includegraphics[width=4cm]{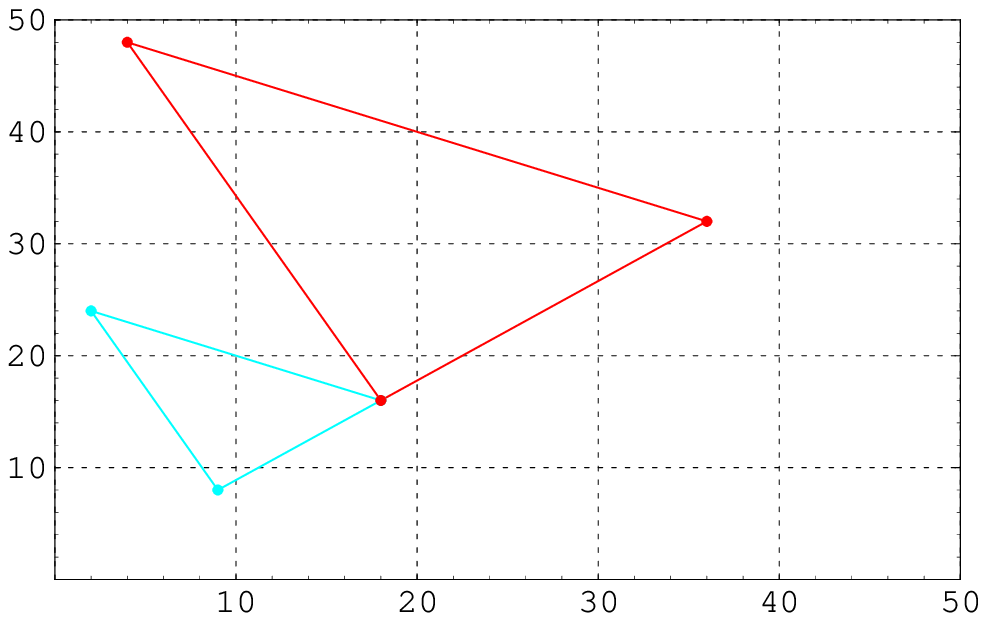}
\includegraphics[width=4cm]{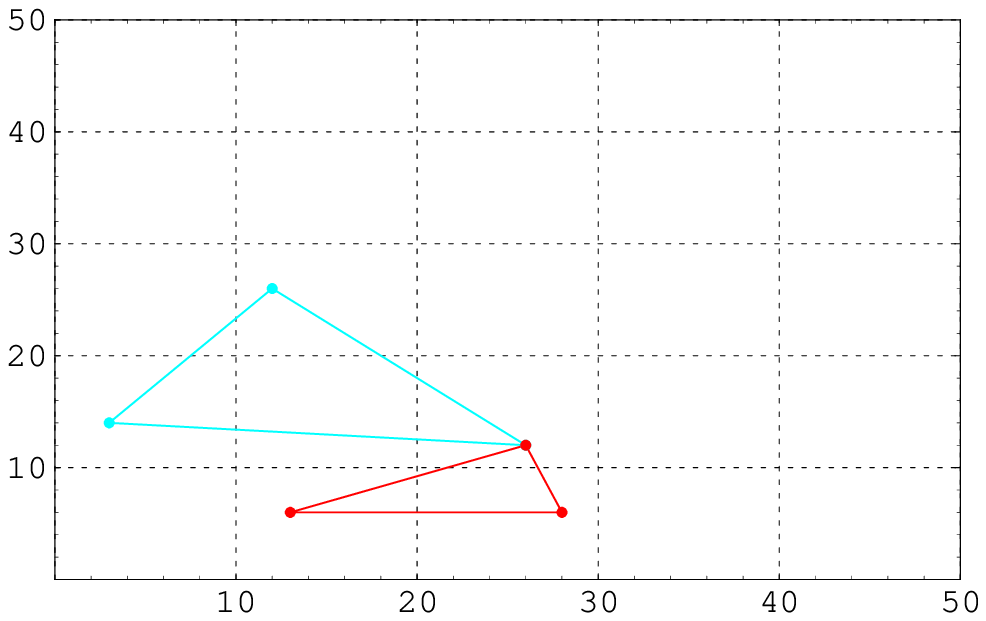}
\end{center}

{\bf 4.} Two further groups consist of two triangles each, but the common point is
contained twice in one triangle (the dynamical systems are isomorphic, but different from the two groups above):
\[
\begin{array}{llll}
 \{& \{\{24,22\},\{32,6\},\{3,16\}\},  & \{\{32,6\},\{32,6\},\{11,12\}\} & \} \\
 \{& \{\{8,38\},\{32,22\},\{11,16\}\}, & \{\{38,8\},\{38,8\},\{11,16\}\} & \}
\end{array}\]
\begin{center}
\includegraphics[width=4cm]{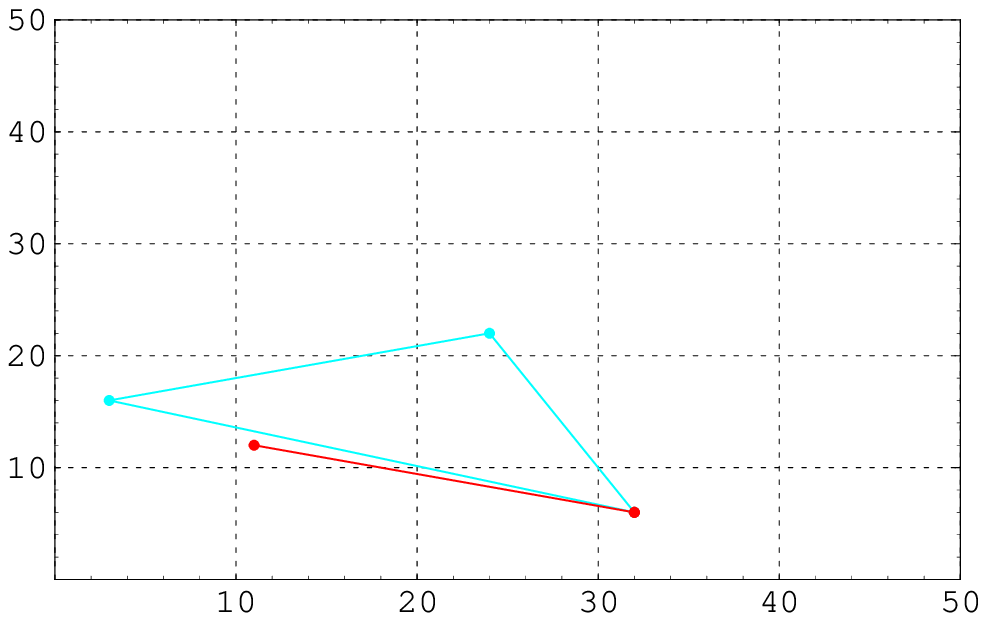}
\includegraphics[width=4cm]{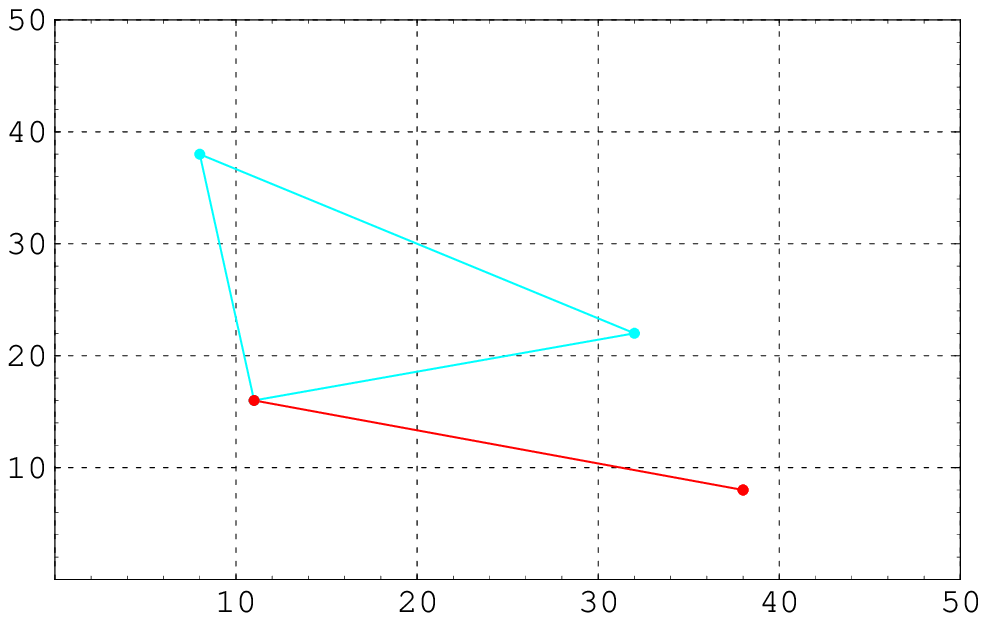}
\end{center}

{\bf 5.} As we can see by inspecting their geometrical structures, further 7 groups are not isomorphic to any group found above:
\[\begin{array}{llll}
\{&
\{\{6,12\},\{12,6\},\{3,6\}\},    & \{\{12,24\},\{24,12\},\{6,12\}\}, \\&
\{\{24,48\},\{48,24\},\{12,24\}\} &
&\}
\end{array}\]
\[\begin{array}{llll}
\{& \{\{2,16\},\{14,8\},\{1,8\}\}, & \{\{4,32\},\{28,16\},\{2,16\}\}, \\&
\{\{32,4\},\{32,4\},\{14,8\}\}   &
&\}
\end{array}\]
\[\begin{array}{llll}
\{&
\{\{2,4\},\{4,2\},\{1,2\}\},&   \{\{4,8\},\{8,4\},\{2,4\}\},  \\&
\{\{8,16\},\{16,8\},\{4,8\}\},& \{\{16,32\},\{32,16\},\{8,16\}\}
&\}
\end{array}\]
\[\begin{array}{llll}
\{&
\{\{4,22\},\{10,20\},\{11,2\}\},  & \{\{8,44\},\{20,40\},\{22,4\}\}, \\&
\{\{10,20\},\{20,10\},\{5,10\}\}, & \{\{20,40\},\{40,20\},\{10,20\}\}
&\}
\end{array}\]
\[\begin{array}{llll}
\{&
\{\{10,40\},\{26,32\},\{19,8\}\},   & \{\{26,32\},\{38,16\},\{13,16\}\}, \\&
 \{\{32,26\},\{40,10\},\{13,16\}\}, & \{\{40,10\},\{40,10\},\{5,20\}\}
&\}
\end{array}\]
\[\begin{array}{llll}
\{&
\{\{4,18\},\{14,12\},\{7,6\}\},   & \{\{8,36\},\{28,24\},\{14,12\}\}, \\&
\{\{12,14\},\{14,12\},\{9,2\}\},  & \{\{24,28\},\{28,24\},\{18,4\}\}, \\&
\{\{36,42\},\{42,36\},\{27,6\}\}, & \{\{42,36\},\{21,18\},\{4,18\}\}
&\}
\end{array}\]
\[\begin{array}{llll}
\{&
\{\{2,36\},\{20,30\},\{17,6\}\},   & \{\{4,6\},\{6,4\},\{3,2\}\}, \\&
\{\{8,12\},\{12,8\},\{6,4\}\},     & \{\{12,18\},\{18,12\},\{9,6\}\}, \\&
\{\{16,24\},\{24,16\},\{12,8\}\},  & \{\{18,12\},\{9,6\},\{4,6\}\}, \\&
\{\{20,30\},\{30,20\},\{15,10\}\}, & \{\{20,30\},\{34,12\},\{1,18\}\}, \\&
\{\{24,36\},\{36,24\},\{18,12\}\}, & \{\{30,20\},\{36,2\},\{1,18\}\}, \\&
\{\{32,48\},\{48,32\},\{24,16\}\}, & \{\{34,12\},\{36,2\},\{15,10\}\}, \\&
\{\{36,24\},\{18,12\},\{8,12\}\},  & \{\{45,30\},\{34,12\},\{12,18\}\}
&\}
\end{array}\]
\begin{center}
\includegraphics[width=4cm]{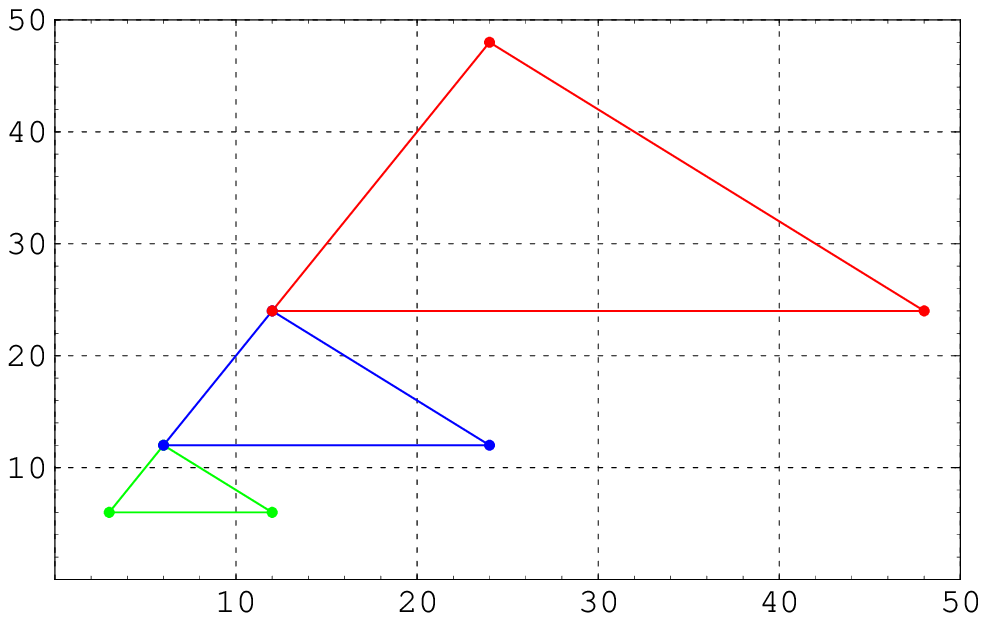}
\includegraphics[width=4cm]{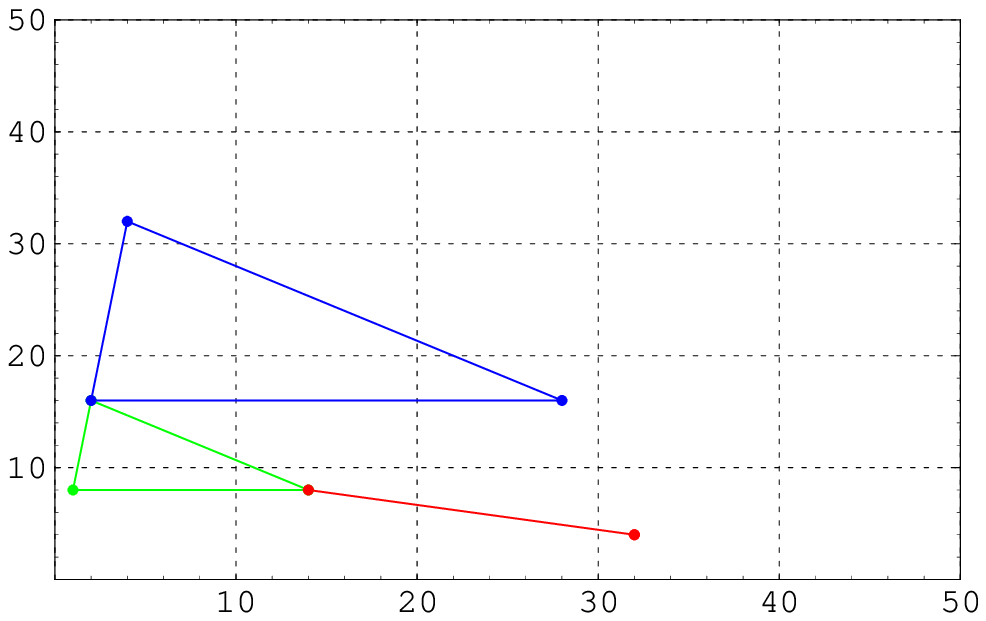}
\includegraphics[width=4cm]{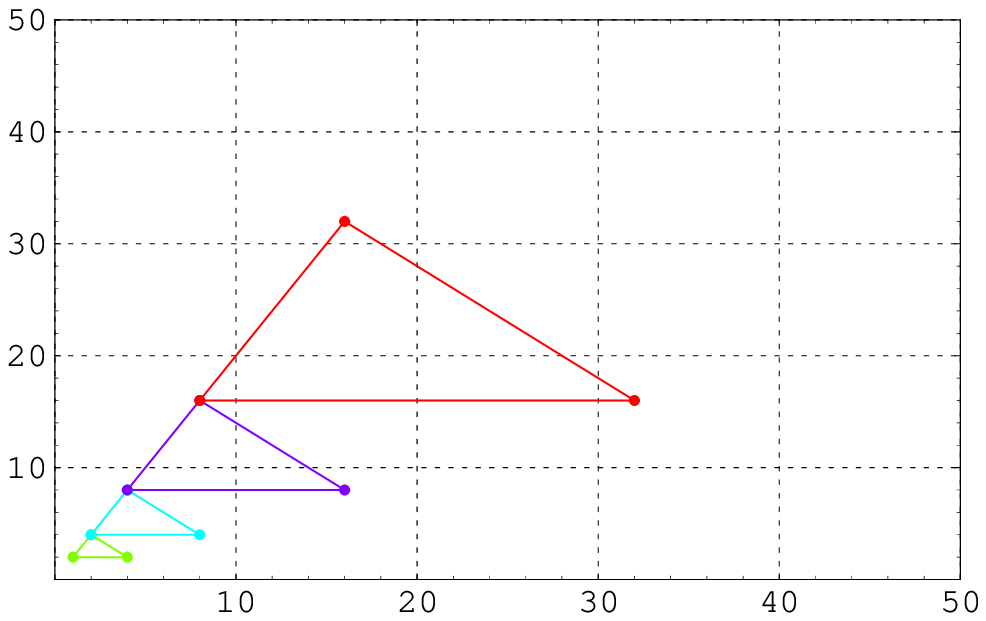}
\includegraphics[width=4cm]{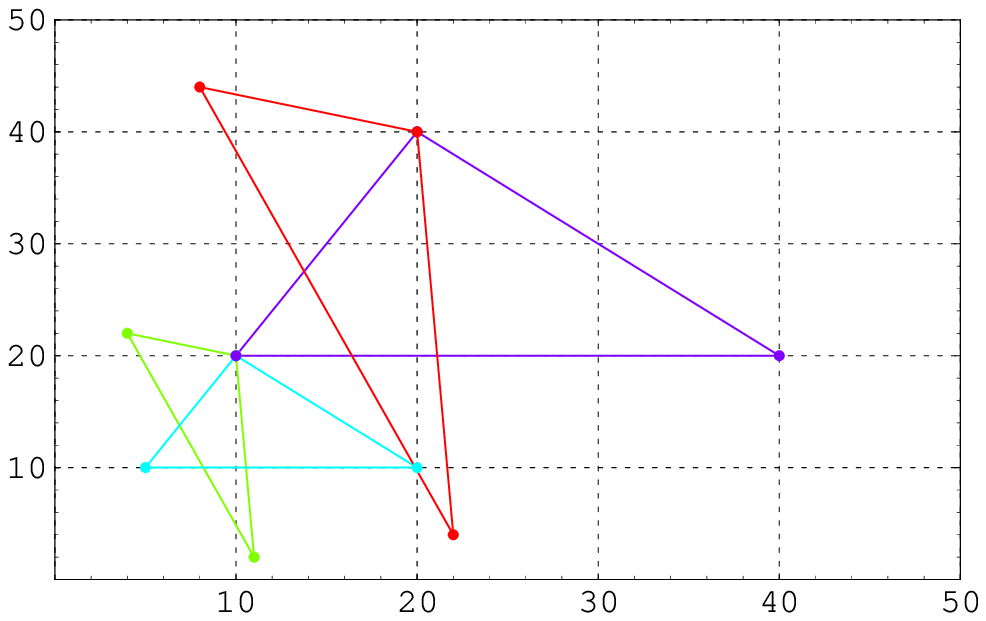}
\includegraphics[width=4cm]{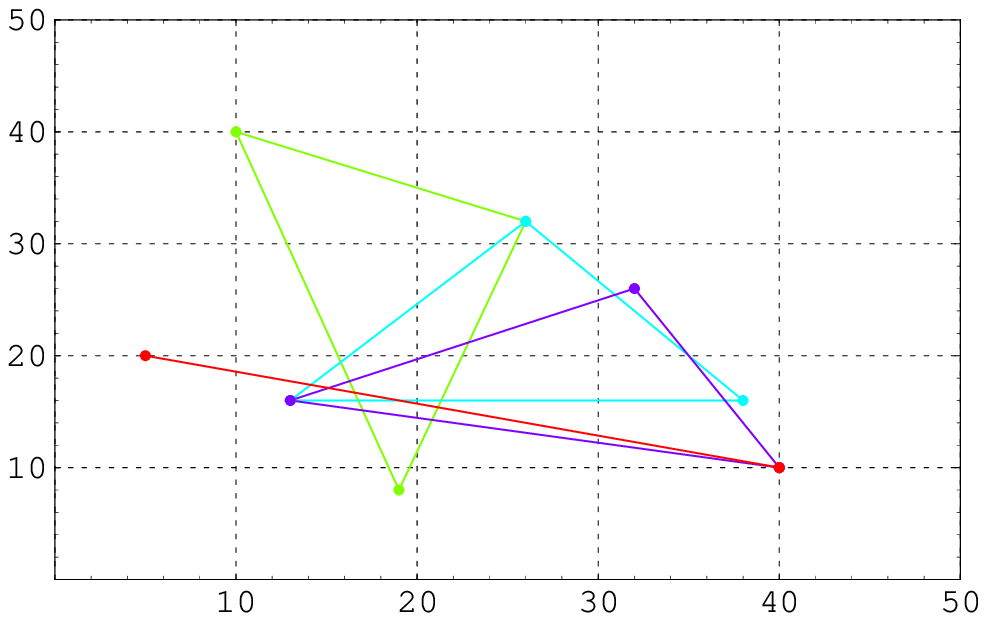}
\includegraphics[width=4cm]{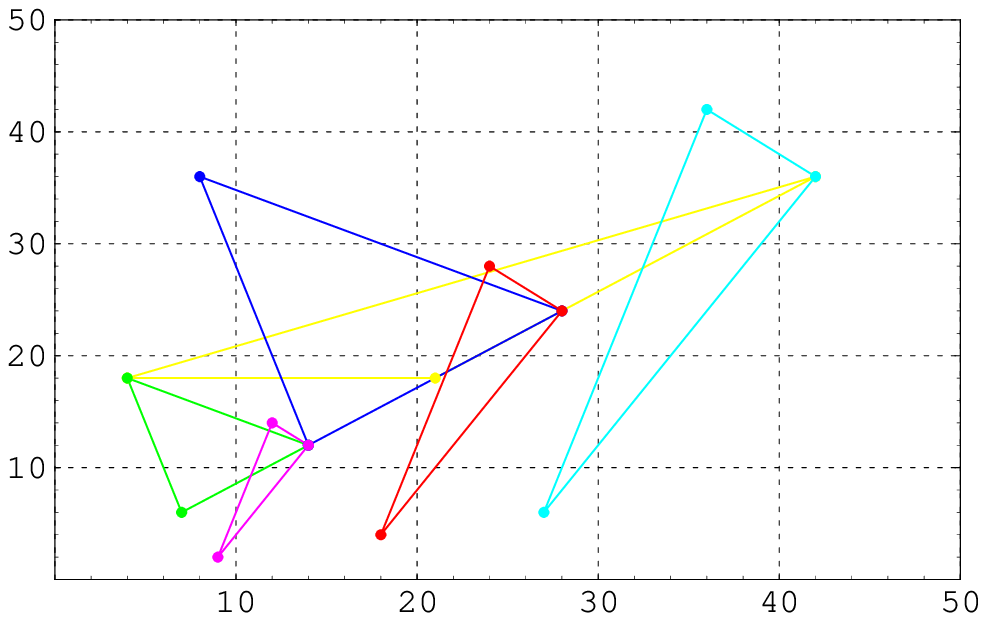}
\includegraphics[width=4cm]{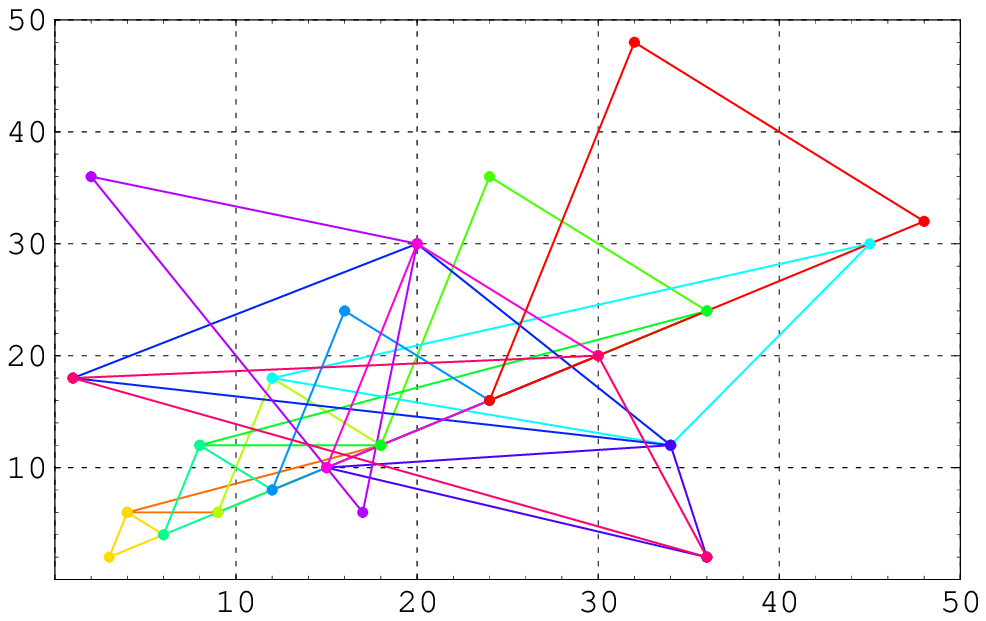}
\end{center}

\subsection{Important Remark}

\begin{figure}[htb]
\begin{center}
        \begin{tabular}{cc}
            \includegraphics[height=3cm]{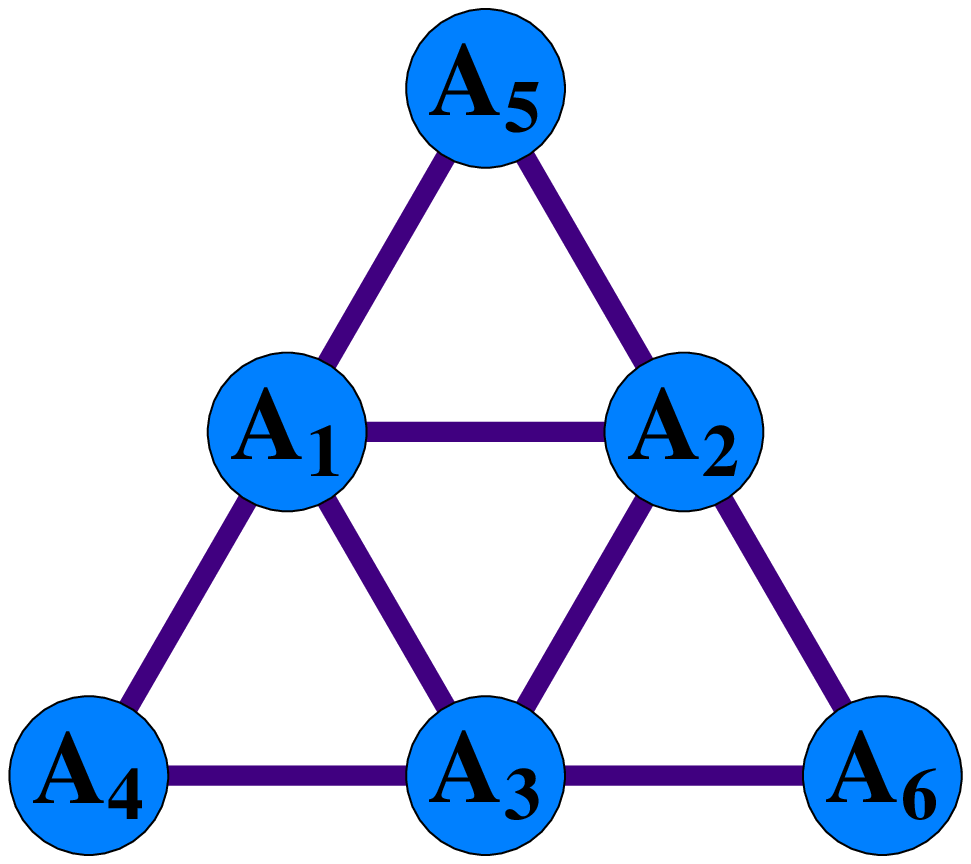} &
            \includegraphics[height=3cm]{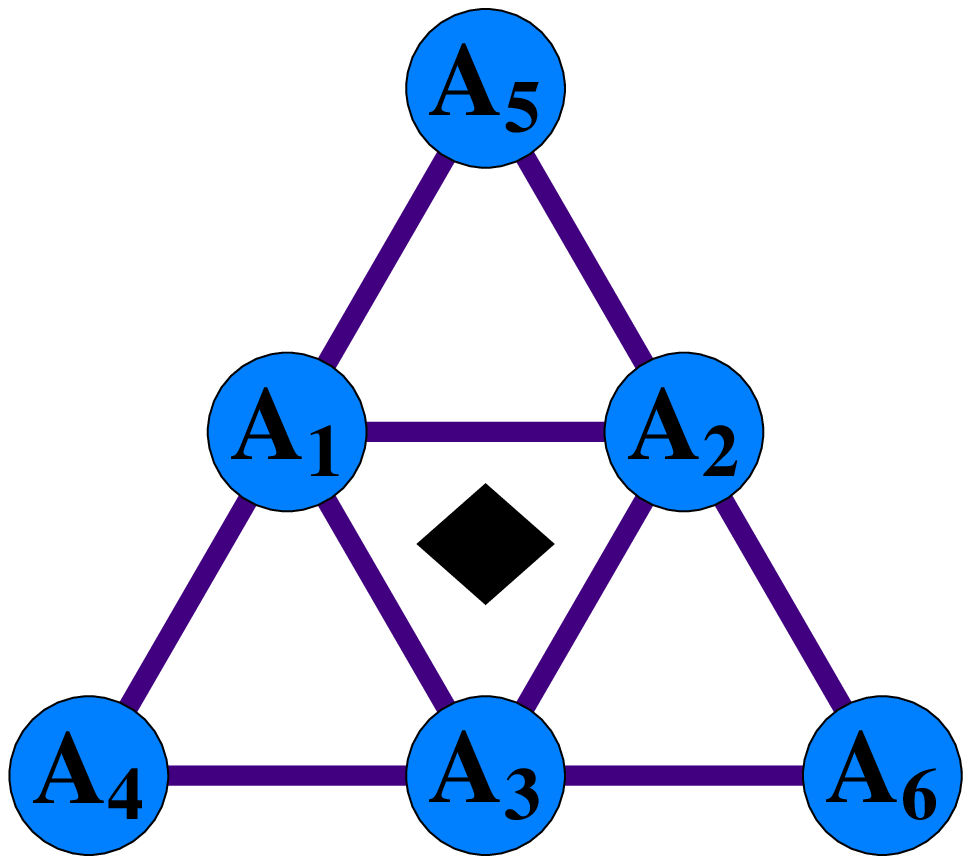} \\
        \end{tabular}
    \caption{Example of isomorphic graphs and
non-isomorphic dynamical systems. The left graph corresponds to the
dynamical system (\ref{dynLeft}) and the graph on the right  - to
the dynamical system (\ref{dynRight}). To discern between these two
cases we set a placeholder inside the triangle not representing a
resonance.}
    \label{fig:HG-Example1}
\end{center}
\end{figure}

To compute all non-isomorphic sub-graphs algorithmically is a
nontrivial problem. Indeed, all isomorphic graphs presented in
previous section are described by similar dynamical systems, only
magnitudes of interaction coefficients $\a_i$ vary. However, in the
general case graph structure thus defined
 does not present the
dynamical system unambiguously. Consider Figure
\ref{fig:HG-Example1} below where two objects are isomorphic {\it as
graphs}. However, the first object represents  4 connected triads
with dynamical system
 \be \label{dynLeft} (A_1,A_2,A_3), \ (A_1,A_2,A_5), \ (A_1,A_3,A_4), \
(A_2,A_3,A_6) \ee
 while the second - 3 connected triads with dynamical system
\be \label{dynRight}  (A_1,A_2,A_5), \ (A_1,A_3,A_4), \
(A_2,A_3,A_6). \ee

This problem has been solved in \cite{KM07} by introducing
hyper-graphs of a special structure; the standard graph isomorphism
algorithm used by Mathematica  has been modified in order to suit
hyper-graphs.

\section{A Web Interface to the Software}


The previous sections have presented implementations of various
symbolic computation methods for the analysis of non-linear wave
resonances. These implementations are written in the language of the
computer algebra system Mathematica which provides an appealing
graphical user interface (GUI) for executing computations and
presenting the results. For instance, the pictures shown in
Section~\ref{results} were produced by converting the computed
hyper-graphs to Mathematica plot structures that can be displayed by
the GUI of the system.

However, to run these methods the user needs an installation of
Mathematica on the local computer with the previously described
methods installed in a local directory. These requirements make
access to the software difficult and hamper its wide-spread usage.
In order to overcome this problem, we have implemented a Web
interface such that the software can be executed from any computer
connected to the Internet via a Web browser without the need for a
local installation of mathematical software.

This implementation follows a general trend in computer science
which turns away from stand alone software (that is installed on
local computers and can be only executed on these computers via a
graphical user interface) and proceeds towards
\emph{service-oriented software}~\cite{service} (that is installed
on remove server computers and wraps each method into a service that
can be invoked over the Internet via standardized Web interfaces).
Various projects in computer mathematics have pursued middleware for
\emph{mathematical web services}, see for instance
\cite{MathBroker,MONET,BaSch}. On the long term, it is thus
envisioned that mathematical methods generally become remote
services that can be invoked by humans (or other software) without
requiring local software installations.

However, even without sophisticated middleware it is nowadays relatively
simple to provide (for restricted application scenarios) web interfaces to
mathematical software by generally available technologies. The web interface
presented in the following sections is deliberately kept as simple as possible
and makes only use of such technologies; thus it should be easy to take this
solution as a blueprint for other mathematical software with similar
features. In particular, the web interface is quite independent of Mathematica
as the system underlying the implementation of the mathematical methods; the
same strategy can be applied to other mathematical software systems such as
Maple, MATLAB, etc.

\subsection{The Interface}

\begin{figure}
\centering\includegraphics[width=0.92\textwidth]{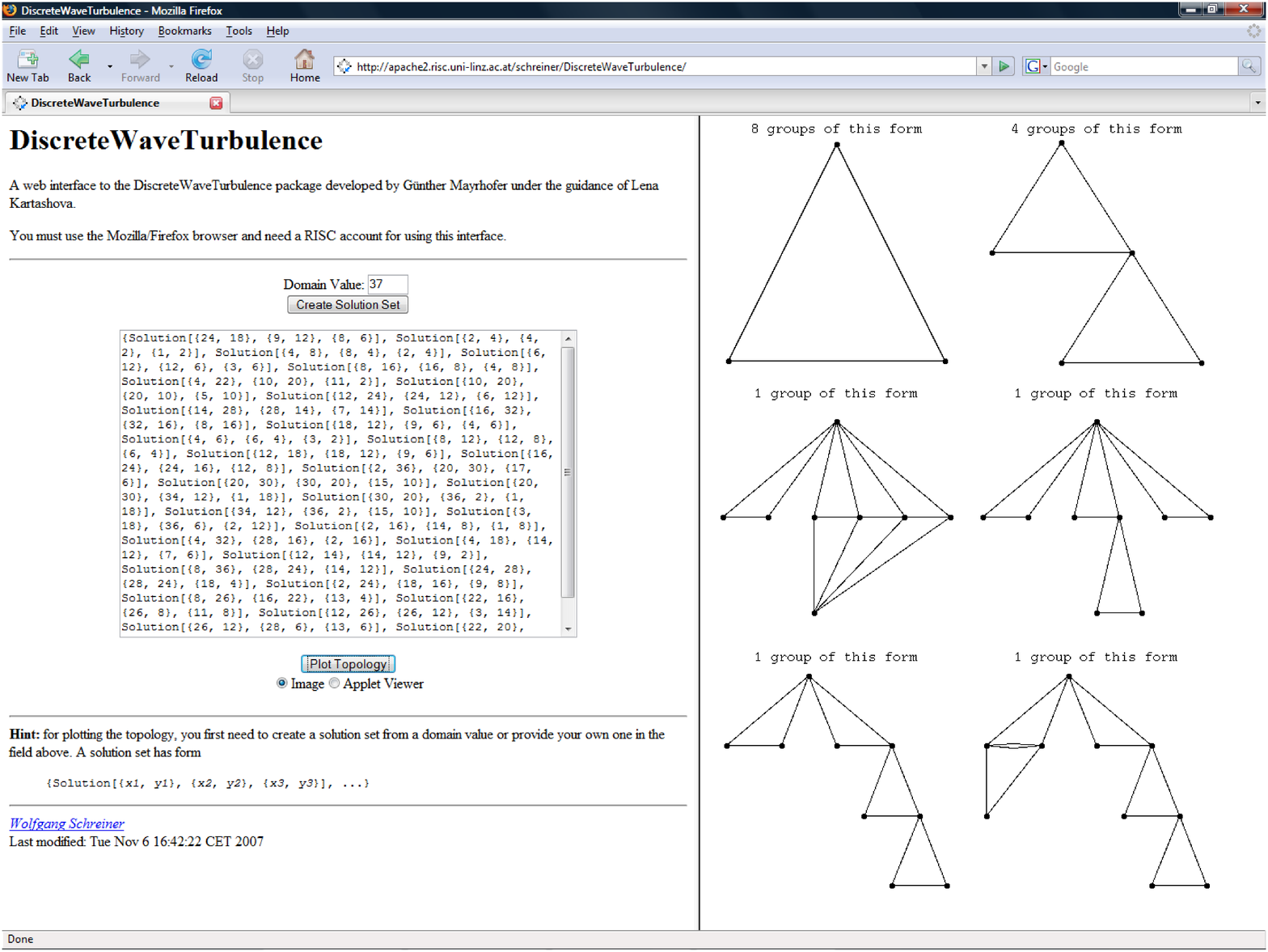}
\caption{Web interface to the implementation}
\label{webinterface}
\end{figure}

Figure~\ref{webinterface} shows the web interface to some of the methods
presented in the previous sections. Its functionality is as follows:
\begin{description}
\item[Create Solution Set] The user may enter a parameter $D$ in the first
  (small) text field and then press the button ``Create Solution Set''. This
  invokes the method \texttt{CreateSolutionSet} which computes the set of all
  solutions whose values are smaller than or equal to $D$. This set is written
  into the second (large) text field in the form
  \begin{quote}
  \texttt{\{Solution[$x_1$,$y_1$,$z_1$],\ldots,Solution[$x_n$,$y_n$,$z_n$]\}}
  \end{quote}
\item[Plot Topology] The user may enter into the second (large) text field a
  specific solution set (or, as show above, compute one), and then press the
  button ``Plot Topology''. This first invokes the method \texttt{Topology}
  which computes the topological structure of the solution set as a list of
  hyper-graphs and then calls the method \texttt{PlotTopology} which
  computes a plot of each hyper-graph. The results are displayed in the right
  frame of the browser window.
\end{description}
The web interface is available at the URL
\begin{center}
\texttt{http://www.risc.uni-linz.ac.at/projects/alisa} \\
(Button ``Discrete Wave Turbulence'')
\end{center}
To run the computations, an account and a password are needed.

\subsection{The Implementation}

\begin{figure}
\begin{center}
\setlength{\unitlength}{4144sp}%
\begingroup\makeatletter\ifx\SetFigFontNFSS\undefined%
\gdef\SetFigFontNFSS#1#2#3#4#5{%
  \reset@font\fontsize{#1}{#2pt}%
  \fontfamily{#3}\fontseries{#4}\fontshape{#5}%
  \selectfont}%
\fi\endgroup%
\begin{picture}(6102,2982)(436,-2773)
\thinlines
{\color[rgb]{0,0,0}\put(4051,-1186){\framebox(2250,900){}}
}%
{\color[rgb]{0,0,0}\put(2251,-61){\line( 0,-1){2700}}
}%
{\color[rgb]{0,0,0}\put(676,-2311){\framebox(1350,1125){}}
}%
{\color[rgb]{0,0,0}\put(1126,-511){\framebox(450,225){}}
}%
{\color[rgb]{0,0,0}\put(766,-826){\framebox(1170,180){}}
}%
{\color[rgb]{0,0,0}\put(1936,-736){\vector( 1, 0){2115}}
}%
{\color[rgb]{0,0,0}\put(3826,-2761){\framebox(2700,2700){}}
}%
{\color[rgb]{0,0,0}\put(4051,-2536){\framebox(2250,900){}}
}%
{\color[rgb]{0,0,0}\put(4951,-1186){\vector( 0,-1){450}}
}%
{\color[rgb]{0,0,0}\put(5401,-1636){\vector( 0, 1){450}}
}%
{\color[rgb]{0,0,0}\put(901,-2626){\framebox(900,180){}}
}%
{\color[rgb]{0,0,0}\put(541,-2716){\dashbox{57}(1620,1710){}}
}%
{\color[rgb]{0,0,0}\put(451,-2761){\framebox(2700,2700){}}
}%
{\color[rgb]{0,0,0}\put(4051,-1096){\vector(-1, 0){1890}}
}%
\put(2296,-196){\makebox(0,0)[lb]{\smash{{\SetFigFontNFSS{8}{9.6}{\ttdefault}{\mddefault}{\updefault}{\color[rgb]{0,0,0}result}%
}}}}
\put(496,-196){\makebox(0,0)[lb]{\smash{{\SetFigFontNFSS{8}{9.6}{\ttdefault}{\mddefault}{\updefault}{\color[rgb]{0,0,0}input}%
}}}}
\put(1351,-1726){\makebox(0,0)[lb]{\smash{{\SetFigFontNFSS{8}{9.6}{\rmdefault}{\mddefault}{\itdefault}{\color[rgb]{0,0,0}S}%
}}}}
\put(5176,-2131){\makebox(0,0)[b]{\smash{{\SetFigFontNFSS{10}{12.0}{\familydefault}{\mddefault}{\updefault}{\color[rgb]{0,0,0}Mathematica}%
}}}}
\put(5131,-871){\makebox(0,0)[b]{\smash{{\SetFigFontNFSS{10}{12.0}{\familydefault}{\mddefault}{\updefault}{\color[rgb]{0,0,0}PHP Engine}%
}}}}
\put(5176,-691){\makebox(0,0)[b]{\smash{{\SetFigFontNFSS{10}{12.0}{\familydefault}{\mddefault}{\updefault}{\color[rgb]{0,0,0}Web Server/}%
}}}}
\put(3826, 74){\makebox(0,0)[lb]{\smash{{\SetFigFontNFSS{10}{12.0}{\familydefault}{\mddefault}{\updefault}{\color[rgb]{0,0,0}Server Computer}%
}}}}
\put(1351,-781){\makebox(0,0)[b]{\smash{{\SetFigFontNFSS{8}{9.6}{\familydefault}{\mddefault}{\updefault}{\color[rgb]{0,0,0}Create Solution Set}%
}}}}
\put(1351,-2581){\makebox(0,0)[b]{\smash{{\SetFigFontNFSS{8}{9.6}{\familydefault}{\mddefault}{\updefault}{\color[rgb]{0,0,0}Plot Topology}%
}}}}
\put(2341,-601){\makebox(0,0)[b]{\smash{{\SetFigFontNFSS{8}{9.6}{\familydefault}{\mddefault}{\updefault}{\color[rgb]{0,0,0}(1)}%
}}}}
\put(3916,-601){\makebox(0,0)[lb]{\smash{{\SetFigFontNFSS{8}{9.6}{\rmdefault}{\mddefault}{\itdefault}{\color[rgb]{0,0,0}D}%
}}}}
\put(451, 74){\makebox(0,0)[lb]{\smash{{\SetFigFontNFSS{10}{12.0}{\familydefault}{\mddefault}{\updefault}{\color[rgb]{0,0,0}Client Computer}%
}}}}
\put(1351,-421){\makebox(0,0)[b]{\smash{{\SetFigFontNFSS{8}{9.6}{\rmdefault}{\mddefault}{\itdefault}{\color[rgb]{0,0,0}D}%
}}}}
\put(3241,-601){\makebox(0,0)[b]{\smash{{\SetFigFontNFSS{8}{9.6}{\ttdefault}{\mddefault}{\updefault}{\color[rgb]{0,0,0}CreateSolutionSet.php/ }%
}}}}
\put(4771,-1456){\makebox(0,0)[b]{\smash{{\SetFigFontNFSS{8}{9.6}{\rmdefault}{\mddefault}{\itdefault}{\color[rgb]{0,0,0}D}%
}}}}
\put(5536,-1456){\makebox(0,0)[b]{\smash{{\SetFigFontNFSS{8}{9.6}{\familydefault}{\mddefault}{\updefault}{\color[rgb]{0,0,0}(3)}%
}}}}
\put(4906,-1456){\makebox(0,0)[rb]{\smash{{\SetFigFontNFSS{8}{9.6}{\ttdefault}{\mddefault}{\updefault}{\color[rgb]{0,0,0}CreateSolutionSet[~~]}%
}}}}
\put(3331,-1456){\makebox(0,0)[b]{\smash{{\SetFigFontNFSS{8}{9.6}{\familydefault}{\mddefault}{\updefault}{\color[rgb]{0,0,0}(2)}%
}}}}
\put(5671,-1456){\makebox(0,0)[lb]{\smash{{\SetFigFontNFSS{8}{9.6}{\rmdefault}{\mddefault}{\itdefault}{\color[rgb]{0,0,0}S}%
}}}}
\put(586,-1141){\makebox(0,0)[lb]{\smash{{\SetFigFontNFSS{8}{9.6}{\ttdefault}{\mddefault}{\updefault}{\color[rgb]{0,0,0}textarea}%
}}}}
\put(2476,-1006){\makebox(0,0)[lb]{\smash{{\SetFigFontNFSS{8}{9.6}{\ttdefault}{\mddefault}{\updefault}{\color[rgb]{0,0,0}<html>..  ..</html>}%
}}}}
\put(2341,-1006){\makebox(0,0)[b]{\smash{{\SetFigFontNFSS{8}{9.6}{\familydefault}{\mddefault}{\updefault}{\color[rgb]{0,0,0}(4)}%
}}}}
\put(3016,-1006){\makebox(0,0)[lb]{\smash{{\SetFigFontNFSS{8}{9.6}{\rmdefault}{\mddefault}{\itdefault}{\color[rgb]{0,0,0}S}%
}}}}
\end{picture}%
 \\ \bigskip\bigskip
\setlength{\unitlength}{0.00087489in}
\begingroup\makeatletter\ifx\SetFigFontNFSS\undefined%
\gdef\SetFigFontNFSS#1#2#3#4#5{%
  \reset@font\fontsize{#1}{#2pt}%
  \fontfamily{#3}\fontseries{#4}\fontshape{#5}%
  \selectfont}%
\fi\endgroup%
{\renewcommand{\dashlinestretch}{30}
\begin{picture}(6102,3897)(0,-10)
\path(3615,3387)(5865,3387)(5865,2487)
	(3615,2487)(3615,3387)
\put(4740,687){\ellipse{900}{180}}
\put(4740,327){\ellipse{900}{180}}
\put(2265,2101){\ellipse{128}{128}}
\path(1815,3612)(1815,912)
\path(240,2487)(1590,2487)(1590,1362)
	(240,1362)(240,2487)
\path(690,3387)(1140,3387)(1140,3162)
	(690,3162)(690,3387)
\path(330,3027)(1500,3027)(1500,2847)
	(330,2847)(330,3027)
\path(3615,2037)(5865,2037)(5865,1137)
	(3615,1137)(3615,2037)
\path(4515,2487)(4515,2037)
\blacken\path(4485.000,2157.000)(4515.000,2037.000)(4545.000,2157.000)(4485.000,2157.000)
\path(4965,2037)(4965,2487)
\blacken\path(4995.000,2367.000)(4965.000,2487.000)(4935.000,2367.000)(4995.000,2367.000)
\path(465,1227)(1365,1227)(1365,1047)
	(465,1047)(465,1227)
\dashline{60.000}(105,2667)(1725,2667)(1725,957)
	(105,957)(105,2667)
\path(15,3612)(2715,3612)(2715,912)
	(15,912)(15,3612)
\path(1365,1137)(1860,1137)(1860,2937)(3615,2937)
\blacken\path(3495.000,2907.000)(3615.000,2937.000)(3495.000,2967.000)(3495.000,2907.000)
\path(4740,1137)(4740,777)
\blacken\path(4710.000,897.000)(4740.000,777.000)(4770.000,897.000)(4710.000,897.000)
\path(4290,687)(4290,327)
\path(5190,687)(5190,327)
\path(3615,2712)(2715,2712)
\blacken\path(2835.000,2742.000)(2715.000,2712.000)(2835.000,2682.000)(2835.000,2742.000)
\path(3390,3612)(6090,3612)(6090,12)
	(3390,12)(3390,3612)
\path(2265,2037)(2265,462)(4290,462)
\blacken\path(4170.000,432.000)(4290.000,462.000)(4170.000,492.000)(4170.000,432.000)
\put(1860,3477){\makebox(0,0)[lb]{\smash{{\SetFigFontNFSS{8}{9.6}{\ttdefault}{\mddefault}{\updefault}result}}}}
\put(60,3477){\makebox(0,0)[lb]{\smash{{\SetFigFontNFSS{8}{9.6}{\ttdefault}{\mddefault}{\updefault}input}}}}
\put(4740,1542){\makebox(0,0)[b]{\smash{{\SetFigFontNFSS{10}{12.0}{\familydefault}{\mddefault}{\updefault}Mathematica}}}}
\put(4695,2802){\makebox(0,0)[b]{\smash{{\SetFigFontNFSS{10}{12.0}{\familydefault}{\mddefault}{\updefault}PHP Engine}}}}
\put(4740,2982){\makebox(0,0)[b]{\smash{{\SetFigFontNFSS{10}{12.0}{\familydefault}{\mddefault}{\updefault}Web Server/}}}}
\put(3390,3747){\makebox(0,0)[lb]{\smash{{\SetFigFontNFSS{10}{12.0}{\familydefault}{\mddefault}{\updefault}Server Computer}}}}
\put(915,2892){\makebox(0,0)[b]{\smash{{\SetFigFontNFSS{8}{9.6}{\familydefault}{\mddefault}{\updefault}Create Solution Set}}}}
\put(915,1092){\makebox(0,0)[b]{\smash{{\SetFigFontNFSS{8}{9.6}{\familydefault}{\mddefault}{\updefault}Plot Topology}}}}
\put(15,3747){\makebox(0,0)[lb]{\smash{{\SetFigFontNFSS{10}{12.0}{\familydefault}{\mddefault}{\updefault}Client Computer}}}}
\put(150,2532){\makebox(0,0)[lb]{\smash{{\SetFigFontNFSS{8}{9.6}{\ttdefault}{\mddefault}{\updefault}textarea}}}}
\put(1860,3027){\makebox(0,0)[lb]{\smash{{\SetFigFontNFSS{8}{9.6}{\rmdefault}{\mddefault}{\updefault}(1)}}}}
\put(915,1947){\makebox(0,0)[b]{\smash{{\SetFigFontNFSS{8}{9.6}{\rmdefault}{\mddefault}{\itdefault}S}}}}
\put(2085,3027){\makebox(0,0)[lb]{\smash{{\SetFigFontNFSS{8}{9.6}{\ttdefault}{\mddefault}{\updefault}PlotTopology.php/}}}}
\put(3255,3027){\makebox(0,0)[b]{\smash{{\SetFigFontNFSS{8}{9.6}{\rmdefault}{\mddefault}{\itdefault}S}}}}
\put(2985,2217){\makebox(0,0)[lb]{\smash{{\SetFigFontNFSS{8}{9.6}{\ttdefault}{\mddefault}{\updefault}PlotTopology[\ldots~~\ldots]}}}}
\put(4110,2217){\makebox(0,0)[b]{\smash{{\SetFigFontNFSS{8}{9.6}{\rmdefault}{\mddefault}{\itdefault}S}}}}
\put(2760,2217){\makebox(0,0)[lb]{\smash{{\SetFigFontNFSS{8}{9.6}{\rmdefault}{\mddefault}{\updefault}(2)}}}}
\put(2760,867){\makebox(0,0)[lb]{\smash{{\SetFigFontNFSS{8}{9.6}{\rmdefault}{\mddefault}{\updefault}(3)}}}}
\put(2985,867){\makebox(0,0)[lb]{\smash{{\SetFigFontNFSS{8}{9.6}{\ttdefault}{\mddefault}{\updefault}Export["image-1.png",\ldots]}}}}
\put(5055,2217){\makebox(0,0)[lb]{\smash{{\SetFigFontNFSS{8}{9.6}{\rmdefault}{\mddefault}{\updefault}(4)}}}}
\put(2130,2532){\makebox(0,0)[lb]{\smash{{\SetFigFontNFSS{8}{9.6}{\ttdefault}{\mddefault}{\updefault}<html><img src="image-1.png">\ldots}}}}
\put(2310,1452){\makebox(0,0)[lb]{\smash{{\SetFigFontNFSS{8}{9.6}{\rmdefault}{\mddefault}{\updefault}(6)}}}}
\put(2535,1452){\makebox(0,0)[lb]{\smash{{\SetFigFontNFSS{8}{9.6}{\ttdefault}{\mddefault}{\updefault}GET image-1.png}}}}
\put(5280,2217){\makebox(0,0)[lb]{\smash{{\SetFigFontNFSS{8}{9.6}{\rmdefault}{\mddefault}{\itdefault}N}}}}
\put(1905,2532){\makebox(0,0)[lb]{\smash{{\SetFigFontNFSS{8}{9.6}{\rmdefault}{\mddefault}{\updefault}(5)}}}}
\end{picture}
} \\
\end{center}
\caption{Implementation of the web interface}
\label{webimpl}
\end{figure}

The web interface is implemented in PHP, a scripting language for producing
dynamic web pages~\cite{PHP}. PHP scripts can be embedded into conventional
HTML pages within tags of form \texttt{<php?\ldots ?>}; when a Web browser
requests such a page, the Web server executes the scripts with the help of an
embedded PHP engine, replaces the tags by the generated output, and returns
the resulting HTML page to the browser. With the use of PHP, thus programs can
be be implemented that run on a web server and deliver their results to a
client computer which displays them in a web browser.  The web interface to
the discrete wave turbulence package is implemented in PHP as sketched in
Figure~\ref{webimpl} and described below (the parenthesized numbers in
the text refer to the corresponding numbers in the figure).

\paragraph{Create Solution Set} The browser frame \texttt{input} on
the left side contains  essentially the following HTML input form:
{\small
\begin{verbatim}
  <form target="textarea"
      action="https://apache2.../CreateSolutionSet.php"
      method="post">
    <input name="domain" size="3">
    <input type="submit" value="Create Solution Set">
 </form>
\end{verbatim}
} This form consists of an input field \texttt{domain} to receive a domain
value and a button to trigger the creation of the solution set. When the
button is pressed, (1) a request is sent to the web server which carries the
value of \texttt{domain}; this request asks the server to deliver the
PHP-enhanced web page \texttt{CreateSolutionSet.php} into the target frame
\texttt{textarea} which is displayed internally to \texttt{input}.

The file \texttt{CreateSolutionSet.php} has essentially the content
{\small
\begin{verbatim}
  <?php
    $math="/.../math";
    $cwd="/.../DiscreteWaveTurbulence";
    $domain = $_POST['domain'];
    $mcmd =
      "SetDirectory[\"" . $cwd . "\"]; " .
      "Needs[\"DiscreteWaveTurbulence`SolutionSet`\"]; " .
      "sol=DiscreteWaveTurbulence`SolutionSet`CreateSolutionSet[" .
         $domain . "]; ";
    $command="$math -noprompt -run '" . $mcmd .
       "Print[StandardForm[sol]]; Quit[];'";
    $result = shell_exec("$command");
    echo
      ...
      "<textarea  name=\"sol\" cols=\"60\" rows=\"20\">" .
      htmlspecialchars($result) .
      "</textarea>" .
     ...;
   ?>
\end{verbatim}
} After setting the paths \texttt{\$math} of the Mathematica binary
and \texttt{\$cwd} of the directory where the
\texttt{DiscreteWaveTurbulence} package is installed, the script
sets the local variable \texttt{\$domain} to the value of the input
field \texttt{domain}. Then the Mathematica command \texttt{\$mcmd}
is constructed in order to load the file \texttt{SolutionSet.m} and execute
the command \texttt{CreateSolutionSet} to compute the solution set.
Now the system command \texttt{\$command} is constructed to (2)
invoke Mathematica which calls the previously constructed command
and (3) prints its result to the standard output stream which is
captured in the variable \texttt{\$result}. From this, the script
contstructs the HTML code of the result document which is (4)
delivered to the Web browser.

\paragraph{Plot Topology} The browser frame \texttt{textarea} contains
essentially the following HTML input form:
{\small
\begin{verbatim}
  <form target="result"
      action="https://apache2..../PlotTopology.php"
      method="post">
    <textarea name="sol" cols="60" rows="20">...</textarea>
    <input type="submit" value="Plot Topology">
    </center>
  </form>
\end{verbatim}
}
This form consists of the textarea field \texttt{sol} to receive the solution
set and a button to trigger the plotting of the topology of this set.  When
the button is pressed, (1) a request is sent to the web server which carries
the value of \texttt{sol}; this request asks the server to deliver the
PHP-enhanced web page \texttt{PlotTopology.php} into the target frame
\texttt{result} on the right side of the browser.

The file \texttt{CreateSolutionSet.php} has essentially the content
{\small
\begin{verbatim}
  <?php
     $math="/.../math";
     $basedir ="/.../DiscreteWaveTurbulence";
     $baseurl ="http://apache2/.../DiscreteWaveTurbulence";
     $sol = $_POST['sol'];
     ... // create under $basedir a unique subdirectory $dir
     $mcmd =
       "SetDirectory[\"$basedir/$dir\"]; " .
       "Needs[\"DiscreteWaveTurbulence`Topology`\"]; " .
       "Needs[\"DiscreteWaveTurbulence`SolutionSet`\"]; " .
       "top=DiscreteWaveTurbulence`Topology`Topology[$sol]; " .
       "plots=DiscreteWaveTurbulence`Topology`PlotTopology1[top];";
     $command="/usr/bin/Xvnc :20 & export DISPLAY=:20; " .
       "export MATHEMATICA_USERBASE=$basedir/.Mathematica; " .
       "$math -run '" . $mcmd .
       "Print[ExportList[plots,\"$image\"]]; Quit[];'";
     $result = shell_exec("$command | tail -n 1");
     for ($i=0;$i<$result;$i++)
       echo "<img src=\"$baseurl/$dir/image-$i.png\"/>";
  ?>
\end{verbatim}
}
For holding the images to be generated later, the script creates a unique
directory \texttt{\$basedir/\$dir} which is served by the web server under the
url \texttt{\$baseurl/\$dir}. The script extracts the solution set
\texttt{\$sol} from the request and sets up the Mathematica command to compute
its topological structure and generate the plots from which ultimately
the image files will be produced.

For this purpose, however, Mathematica needs an X11 display server
running; since a Web server has not access to an X11 server, we
start the virtual X11 server Xvnc~\cite{Xvnc} as a replacement and
set the environment variable \texttt{DISPLAY} to the display number
on which the number listens; Mathematica will subsequently send X11
requests to that display which will be handled by the virtual
server. Likewise, Mathematica needs access to a
\texttt{.Mathematica} configuration directory; the script sets the
environment variable \texttt{MATHEMATICA\_USERBASE} correspondingly.

With these provisions, we can (2) invoke first the command to
compute the plots and then the (self-defined) command
\texttt{ExportList} to generate for every plot an image
in the previously created directory. For this purpose the command
uses (3) the Mathematica command
\texttt{EXPORT[\emph{file},\emph{plot},"PNG"]} which converts
\emph{plot} to an image in PNG format and writes the image to
\emph{file}. \texttt{ExportList} returns the number of images
generated which is (4) written to the standard output stream which
in turn is captured in the variable \texttt{\$result}. From this
information, the script generates an HTML document which contains a
sequence of \texttt{img} elements referencing these images. After
this document has been (5) returned to the client browser, the
browser (6) requests the referenced images with \texttt{GET}
messages from the web server.

\subsection{Extensions}

As an alternative to the display of static images, the Web interface
also provides an option ``Applet Viewer'' with somewhat more
flexibility. If this option is selected, Mathematica is instructed
to save all generated plots as files in the standard representation.
The generated HTML document then embeds (rather than \texttt{img}
elements) a sequence of \texttt{applet} elements that load instances
of the ``JavaView'' applet~\cite{JavaView}. These applets run in the
Java Virtual Machine of the Web browser on the client computer, load
the plot files from the web, and visualize them in the browser.
Rather than just displaying static images, the viewer allows to
perform certain manipulations and transformations of the plots such
as scaling, rotating, etc. While this additional flexibility is not
of particular importance for the presented methods, they may in the
future become useful for others.

To limit access to the software respectively to the computing power of the
server computer, it may be protected by authentication mechanisms. For
example, on the Apache Web server, it suffices to provide in the installation
directory of the software a file \texttt{.htaccess} with content {\small
\begin{verbatim}
  <Files "*.php">
    SSLRequireSSL
    AuthName "your account"
    AuthType Basic
    Require valid-user
  </Files>
\end{verbatim}
}
With this configuration, the user is asked for the data of a valid account on
the computer running the Web server; other authentication mechanisms based
e.g.\ on password files may be provided in a similar fashion.

\section{Discussion}

Summing up all the results obtained, we would like to make some
concluding remarks.

\begin{itemize}
\item{} In general, coefficients $\a_i$ can be computed symbolically by
hand and only numerically by Mathematica (see Section 3.3); at
present we are not aware of the possibility to overcome this
problem.
\item{} For the known case of spherical barotropic vorticity equation,
values of coefficients $\a_i$ coincide with known form the
literature for all triads but three. These 3 triads, though
satisfying resonant conditions, are known to be special from the
physical point of view  in the following sense (see \cite{KL-06} for
details). Though resonance conditions are fulfilled for the waves of
these triads, they, so to say, do not have a place in the physical
space to interact and their influence (if any) on the dynamics of
the wave system has to be studied separately from all other waves.
Our results might indicate that also the coefficients $\a_i$ of
these triads have to be defined in some other way compare to other
resonant triads. For instance, another way of space-averaging has to
be chosen.
\item{} The results of Section 3.4.2 show that analytical formulae given in \cite{rectangular}
for $\a_i$ {\it are not correct}.
\item{} The results of Section 4.3 show a crucial dependence of the number of solutions on the
form of the boundary conditions. In particular, some boundary
conditions (for example, $\ (L_x,L_y)=(11,29)$) yield {\it no
solutions} which is of most importance for physical applications.
From the mathematical point of view, an interesting result has been
observed: in all our computations (i.e. for $\ m, n \le 300$)
indexes corresponding to non-empty classes, turned out to be {\it
odd.} It would be interesting to prove this fact analytically
because if it keeps true, we can reduce the computational time.
\item{} The algorithm presented in Section 4 has been implemented before numerically
in Visual Basic, and our purpose here was to show that it works fast
enough also in Mathematica. The algorithms presented in Section 3
and Section 5 {\it have never been implemented before}, the whole
work is usually done by hand and some mistakes as in
\cite{rectangular} are almost unavoidable: it takes sometimes a few
weeks of skillful researchers to compute interaction coefficients of
dynamical systems for one specific wave system.
\item{} All the algorithms presented above can  easily be modified for the
case of a 4-term mesoscopic system. The only problem left is a
procedure to establish all non-isomorphic topological elements for a
quadruple graphs, similar to the procedure given in \cite{KM07} for
a triangle graphs. The structure of quadruple graphs is much more
complicated while some mechanisms of energy transfer in the spectral
space do exist \cite{K07} that are absent in 3-term mesoscopic
systems. A complete classification of quadruple graphs is still an
open question but in a given spectral domain it can be done directly
(a very time consuming operation).

\item{} We have developed a Web interface for the presented methods, which
  turns the implementations from only locally available software to Web-based
  services that can be accessed from any computer in the Internet that is
  equipped with a Web browser.  The presented implementation strategy is
  simple and based on generally available technologies; it can be applied as a
  blueprint for a large variety of mathematical software.  In particular, the
  results are not bound to the current Mathematica implementation but can be
  adapted to any other computer algebra system (e.g.\ Maple) or
  numerical
  software system (e.g.\ MATLAB) of similar expressiveness.

\item{} At present, an explicit form of eigen-modes (\ref{spherical_mode}), (\ref{rectangular_mode}) is used
as one of the input parameters for our program package.
Theoretically, at least for some classes of linear partial
differential operators and boundary conditions, computing
eigen-modes can also be performed symbolically basing on the results
in \cite{Markus}. If this were done, not an eigen-mode but boundary
conditions would play role of input parameter.

\end{itemize}

{\bf Acknowledgements.} Authors acknowledge the support of the
Austrian Science Foundation (FWF) under projects SFB F013/F1301
"Numerical and symbolical scientific computing", P20164-N18
"Discrete resonances in nonlinear wave systems" and P17643-NO4
"MathBroker II: Brokering Distributed Mathematical Services".

\end{document}